%% file: MagneticTorque.tex
\documentclass[aps,twocolumn,amsmath,amssymb,a4paper,9pt,showkeys,notitlepage]{revtex4}
\def\pdftitle{Magnetic Torque of Microfabricated Elements and
  Magnetotactic Bacteria}
\def\authorname{Leon Abelmann}
\def\pdfsubject{}
\def\pdfkeywords{}
\def\pdfbackref{none}

\include{includes/LeonHeader}
\include{includes/LeonMacro}

\usepackage[paper=a4paper,total={170mm,241mm},top=20mm,left=20mm,columnsep=8mm]{geometry}\usepackage{siunitx}
\usepackage{nicefrac}

\def\figurewidth{0.8\columnwidth}
\def\widefigurewidth{\columnwidth}

\newif\ifcmtr
\cmtrfalse
\usepackage{color}
\ifcmtr %  Highlight comment (and hide it when necessary)
\newcommand{\cmtr}[1]{ %
   [\color{red} \textbf{#1} \normalcolor]%
}%
\else
\newcommand{\cmtr}[1]{ %
}%
\fi

\def\CoNi{Co$_\text{80}$Ni$_\text{20}$}%

\begin{document}

\title{Magnetic torque on microfabricated elements and magnetotactic bacteria}
\author{Lars Zondervan$^1$}
\author{\"Ozlem Sardan Sukas$^1$}
\author{Islam S. M. Khalil$^{2}$}
\author{Marc P. Pichel$^{3,4}$}
%\author{Hans Kolk$^4$} 
\author{Sarthak Misra$^4$} 
\author{Leon Abelmann$^{1,3}$} 
\affiliation{$^1$MESA$^+$ Research Institute, University of
 Twente, The Netherlands\\$^2$The German University in Cairo, New
 Cairo City, Egypt\\$^{3}$KIST Europe,
 Saarbr\"ucken, Germany\\$^4$MIRA, University of Twente, The Netherlands \\\underline{l.abelmann@kist-europe.de}}
%\date{\today}
\vskip10mm
\begin{abstract}
  \textbf{Abstract:} We present a thorough theoretical analysis of the
  magnetic torque on microfabricated elements with dimensions in the
  range of 100 to \SI{500}{\micro m} and magneto-somes of
  magnetotactic bacteria of a few \si{\micro m} length. We derive
  simple equations for field dependent torque and magnetic shape
  anisotropy that can be readily used to replace the crude
  approximations commonly used. We illustrate and verify the theory on
  microfabricated elements and magnetotactic bacteria, by field
  depedent torque magnetometry and by observing their rotation in
  water under application of a rotating magnetic field. The maximum
  rotation frequency of the largest microfabricated elements agrees
  within error boundaries with theory. For smaller, and especially
  thinner, elements the measured frequencies are a factor of three to
  four too low. We suspect this is caused by incomplete saturation of
  the magnetisation in the elements, which is not incorporated in our
  model. The maximum rotation frequency of magnetotactic bacteria
  agrees with our model within error margins, which are however quite
  big due to the large spread in bacteria morphology. The model
  presented provides a solid basis for the analysis of experiments
  with magnetic objects in liquid, which is for instance the case in
  the field of medical microrobotics.
\end{abstract}

% insert suggested PACS numbers in braces on next line
% insert suggested keywords - APS authors don't need to do this

\maketitle
% \makeatletter 
% \renewcommand{\@tocrmarg}{2.55em plus1fil} 
% \renewcommand{\@pnumwidth}{3em} 
% \renewcommand{\@tocrmarg}{4em} 
% \makeatother

%\tableofcontents

\section{Introduction}

%Zeg hier iets over het gebruik van mxB, compleet negeren van field
%dependence.
% Itt Erglis veel hogere frequenties voor MTB.

Untethered micro-robotic systems for medical applications make a
compelling research field~\cite{Menciassi2007,Abbott2009}.  It is a
form of Minimally Invasive Surgery (MIS), in which one tries to reduce
patient surgical trauma while enabling clinicians to reach deep seated
locations within the human body~\cite{Nelson2004, Abayazid2013}. The
current approach is to insert miniaturised tools needed for a medical
procedure into the patient through a small insertion or orifice. By
reducing the size of these tools a larger range of natural pathways
become available. Currently, these tools are however mechanically
connected to the outside world. If this connection can be removed, so
that the tools become untethered, (autonomous) manoeuvring through
veins and arteries of the body becomes possible~\cite{Dankelman2011}.

If the size and/or application of these untethered systems inside the
human body prohibits storage of energy for propulsion, the energy
has to be harvested from the environment. One solution is the
use of alternating magnetic fields~\cite{Abbott2009}. This method is
simple, but although impressive progress has been made, it is appallingly
inefficient. Only a fraction of~\num{e-12} of the supplied energy
field is actually used by the microrobot.

Efficiency would increase dramatically if the microrobot could harvest
its energy from the surrounding liquid. In human blood, energy is
abundant and used by all cells for respiration. Despite impressive
first attempts~\cite{Pilarek2011}, we are however far away from using
nutrient such as glucose for micro-actuation. As model systems, one
could use simpler fluids to derive energy, such as
peroxide~\cite{Ismagilov2002,Fournier-Bidoz2005, Sundarajan2008,
  Solovev2012}. But of course, peroxide is not compatible with medical
applications.

Nature provides us however with a plentitude of self-propelling
micro-organisms that derive their energy from bio-compatible
liquids. There are even magnetotactic bacteria~\cite{Blakemore1979},
that use the earth magnetic field to locate the bottom of marshes or
ponds. These bacteria are perfect model systems to test concepts and
study the behaviour of self-propelling micro-objects steered by
external magnetic fields~\cite{Khalil2013b}.

For self-propelled objects, only the direction of motion needs to be
controlled by the external magnetic field. There is no need for field
gradients to apply forces, so the field can be uniform. Compared to
systems that derive their energy for propulsion from the magnetic
field, the field can be small in magnitude and needs to vary only
slowly. As a result, the energy requirements are low and overheating is
no longer a problem.

It is generally accepted that magnetotactic bacteria react on the
external magnetic field in a passive fashion, much like compas
needles~\cite{Erglis2007}. The direction of the motion is changed by
application of a magnetic field under an angle with the easy axis of
magnetisation of the micro-objects. The resulting magnetic torque
causes a rotation of the micro-object at a speed that is determined by
the balance between magnetic torque and rotational drag torque. 

The magnetic torque is often modelled by assuming that the magnetic
element is a permanent magnet with dipole moment $\bs{m}$ [Am$^2$] on
which the magnetic field $\bs{B}$ [T] exerts a torque
$\Gamma=\bs{m}\times\bs{B}$ [N]. This simple model suggest that the
torque increases linearly with field, because it is assumed that the
atomic dipoles are rigidly fixed to the lattice, and cannot rotate at
all. This is usually only the case for very small magnetic fields. In
general one should consider changes in magnetic energy as a function
of magnetisation direction with respect to the object (magnetic
anisotropy). This is correctly suggested by Erglis and co-authors for
magnetotactic bacteria, but they fail to analyse the
consequences~\cite{Erglis2007}. Moreover, they suggest that the
magnetic element of the magnetotactic bacterium (the magnetosome),
can be approximated with a solid cilinder, which is a very crude
approximation.

Since magnetotactic bacteria suffer from a large spread in dimensions
and magnetic properties, these approximations are difficult to
verify. Therefore, we have fabricated microlelements of which the
dimensions and magnetic properties are exactly known. We present a
thorough theoretical analysis of the magnetic torque on
microfabricated elements and magnetosomes of magnetotactic
bacteria. We derive simple equations for field dependent torque and
magnetic shape anisotropy that can be readily used to replace the
crude approximations commonly used. We illustrate and verify the
theory on microfabricated elements and magnetic bacteria, by torque
magnetometry and by observing their rotation in water under
application of a rotating magnetic field.

\section{Theory}
\label{sec:theory}
The magnetic torque $\Gamma$ [Nm] is equal to the change in total
magnetic energy $U$ [J] with changing applied field angle. Since our
microfabricated elements and magnetotactic bacteria have negligible
crystal anisotropy, we only consider the demagnetisation and
external field energy terms. The demagnetisation energy is caused by
the magnetic stray field $\bs{H}_\text{d}$ [A/m] that arises due to the
sample magnetisation $\bs{M}$ [A/m] and is mathematically equivalent to~\cite{Hubert1998}

\begin{equation}
U_\text{d}=\frac{1}{2}\mu_0\int\bs{M}\cdot\bs{H_\text{d}}dV
\end{equation}

The demagnetisation energy acts to orient the magnetisation such that
the external stray field energy is minimised. We can define a shape
anisotropy term $K$ [J/m$^{-3}$] to represent the energy difference
between the hard and easy axis of magnetisation, which are 
perpendicular to each other

\begin{equation}
K=\left(U_\text{d, max}-U_\text{d, min}\right)/V
\end{equation}

The external field energy is caused by the externally applied field $\bs{H}$ [A/m]

\begin{equation}
U_\text{H}=-\mu_0\int\bs{M}\cdot\bs{H}dV
\end{equation}

and acts to align $\bs{M}$ parallel to $\bs{H}$. Assuming that the
magnetic element of volume $V$ is uniformly magnetised with saturation
magnetisation $M_\text{s}$ [A/m], the total energy can be thus
expressed as

\begin{equation}
  U=KVsin^2(\theta)-\mu_0M_\text{s}HVcos(\varphi-\theta)
\end{equation}

The angles are defined as in figure~\ref{fig:AxisMandH}a. Normalising
energy, field and torque

\begin{align}
u&=\nicefrac{U}{KV}\\
h&=\nicefrac{\mu_0HM}{2K}\\
\gamma&=\nicefrac{\Gamma}{KV}
\end{align}

the energy expression can be simplified: 

\begin{equation}
u=\sin^2(\theta)-2h\cos(\varphi-\theta)
\end{equation}

The equilibrium magnetisation direction is reached for
$\nicefrac{\partial u}{\partial \theta}$=0, which leads to an
equation that cannot be expressed in an analytically concise
form. The main results are however that for
$h<\nicefrac{1}{\sqrt{2}}$, the maximum torque is reached at the field angle
$\varphi_\text{max}=\nicefrac{\pi}{2}$,

\begin{align}
 \gamma_\text{max}&=2h\sqrt{1-h^2} &\text{ for } h\le\nicefrac{1}{\sqrt{2}}\\
  &=1 &\text{ for } h>\nicefrac{1}{\sqrt{2}}
\end{align}

The angle of magnetisation at maximum torque can be approximated by

\begin{equation}
\theta_\text{max} =h+0.1 h^2  \text{ for
} h< \nicefrac{1}{\sqrt{2}},\\
\end{equation}

where the error is smaller than $5\times10^{-3}$ rad (\SI{1.6}{\degree}) for $h<0.5$.

For $h>1$, the field angle $\varphi_\text{max}$ at which the maximum
torque is reached is smaller than $\nicefrac{\pi}{2}$ and approaches
$\nicefrac{\pi}{4}$ for $h\rightarrow\infty$. This behaviour can be very
well approximated by

\begin{align}
\varphi_\text{max}=\frac{\pi}{4}\left(1+\frac{2}{3h}\right) \text{ for
} h>1,
\end{align}

 where the error is smaller than $3\times10^{-3}\pi$
(\SI{0.5}{\degree}).

In summary, and returning to variables with units, the maximum
torque is $\Gamma_\text{max}=KV $ which is reached at 

\begin{equation}
H>\frac{\sqrt{2}K}{\mu_0M_\text{s}}
\end{equation}

at an angle $\varphi=\nicefrac{\pi}{2}$, which decreases in good approximation
linearly with $1/H$ to
$\varphi=\nicefrac{\pi}{4}$ at infinite external field. In the
following, we determine the value of the anisotropy energy density $K$
for microfabricated elements as well as magnetotactic bacteria.

\subsection{Magnetic torque of microfabricated elements}
The microfabricated elements are thin film rectangular elements. We
can calculate the magnetic torque either for field applied
perpendicular or in the film plane.

\subsubsection{Perpendicular fields}
\label{sec:perpendicular-fields}
\begin{figure}
  \begin{centering}
    \subfloat[Out-of-plane anisotropy]{
      \begin{centering}
        \includegraphics[width=\figurewidth]{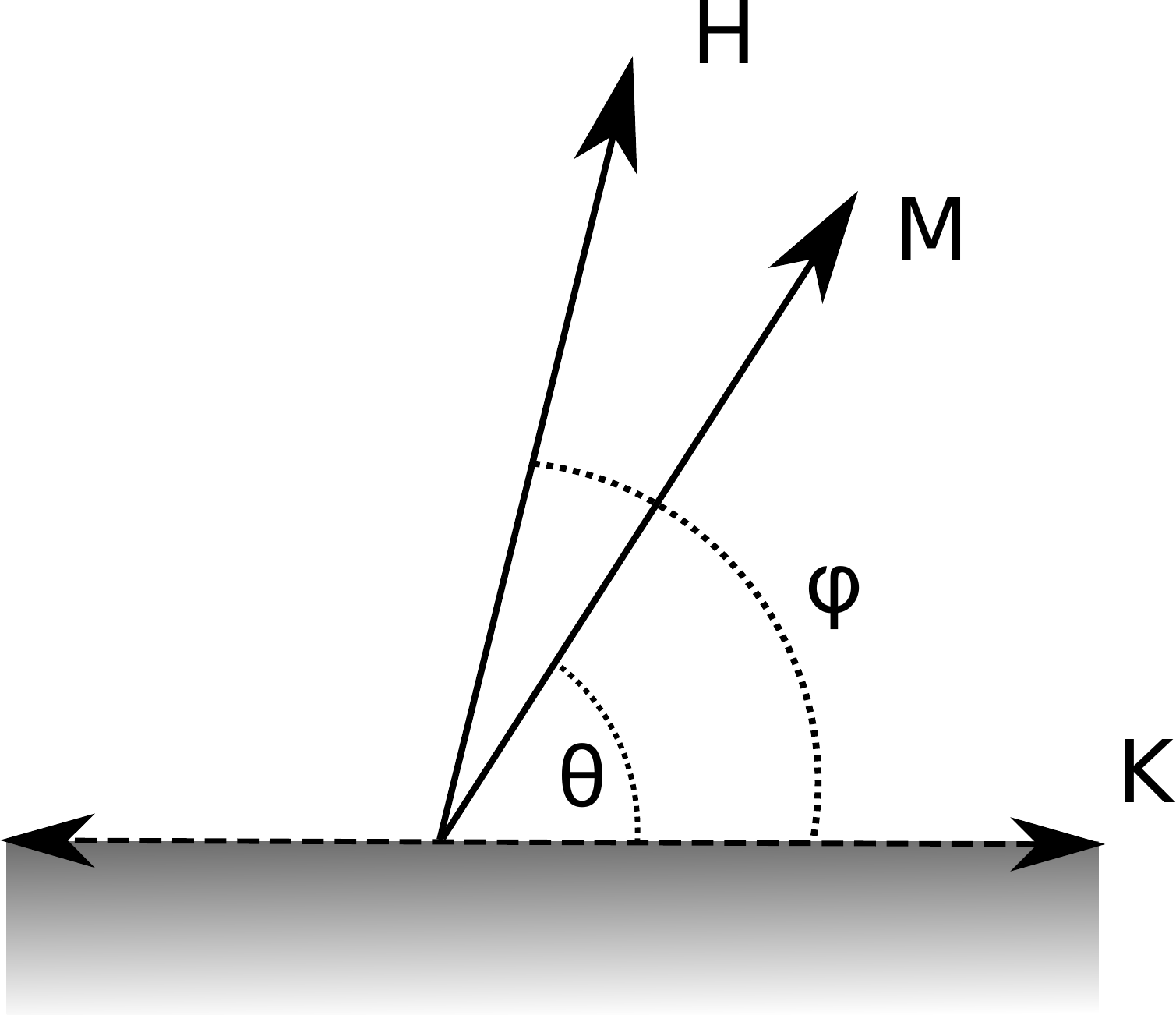}
      \end{centering} }
  \end{centering}
  \begin{centering}
    \subfloat[In-plane anisotropy]{
      \begin{centering}
        \includegraphics[width=\figurewidth]{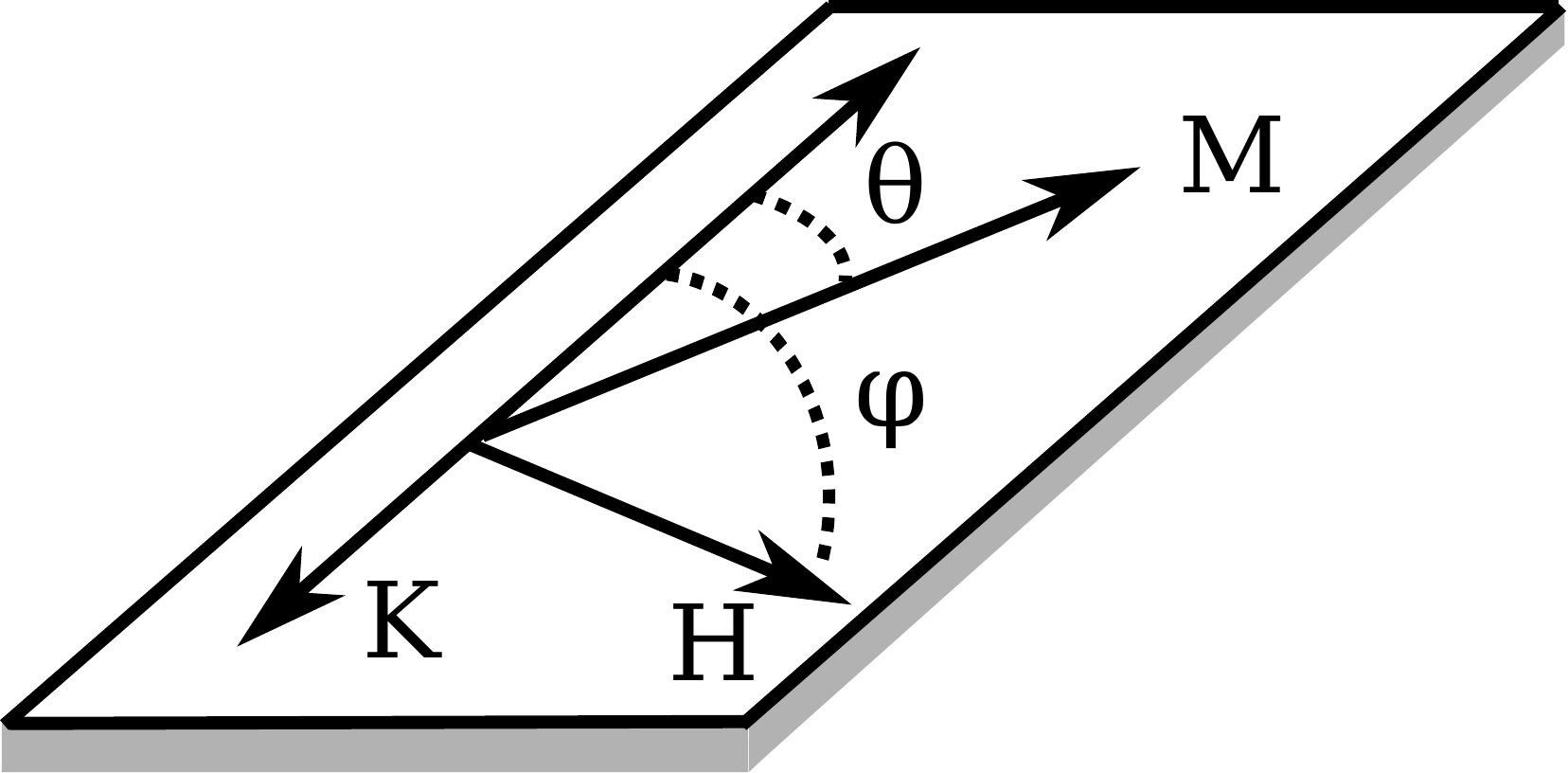}
      \end{centering} }
  \end{centering}
  \caption{Definition of the field angle $\varphi$ and the magnetisation angle
    $\theta$ between the easy axis $K$ of the thin film, the magnetization
    $M$ and the magnetic field $H$ for (a) out-of-plane and
    (b) in-plane anisotropy.}
  \label{fig:AxisMandH}
\end{figure}

The microfabricated elements have lateral dimensions much larger than
the film thickness. Therefore, the demagnetisation field strength is zero for
fields applied in the film plane and equal to $M$ if the field is
directed perpendicular to the film plane.  The maximum demagnetisation
energy difference for fields applied in a plane perpendicular to the
film plane (figure~\ref{fig:AxisMandH}a), is simply

\begin{equation}
KV=\Gamma_{\perp\text{max}}=\frac{1}{2}\mu_0M_\text{s}^2V
\end{equation}

\subsubsection{In-plane fields}

For fields applied in the thin film plane, the shape energy term is
more complex. If again we assume that the external field is
sufficiently high, so that the magnetisation aligns with the field, the
maximum torque is determined by the difference in magnetic energy for
the field aligned along either edge of the film.

\begin{figure}
  \begin{centering}
    \includegraphics[width=\figurewidth]{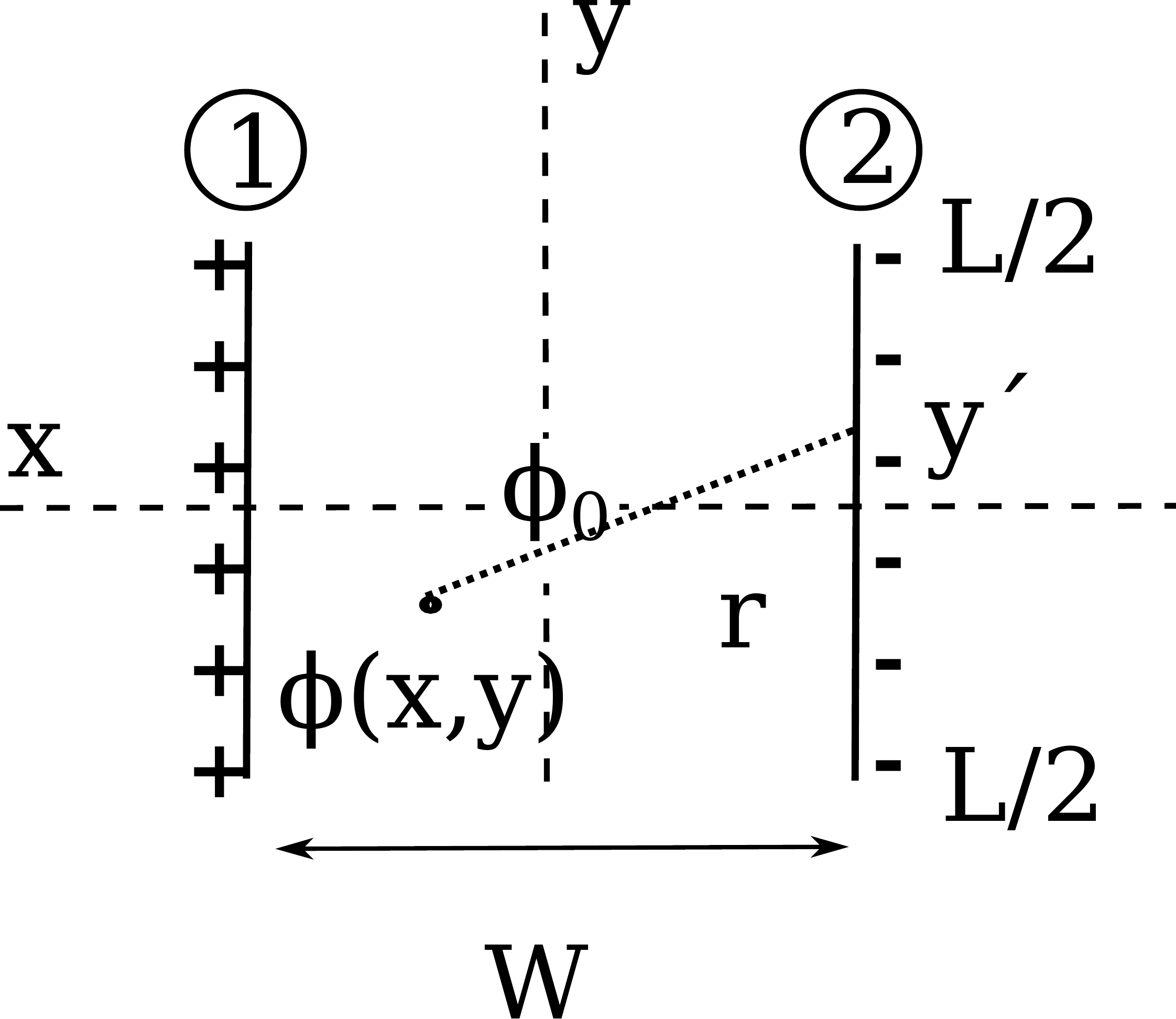}
  \end{centering}
  \caption{Charged line approximation of an in-plane thin film magnetised along
    its width $W$.}
  \label{fig:inplane}
\end{figure}

Since there are no free currents, we can define a magnetic scalar
potential $\phi$ and consider magnetic
charges~\cite{FeynmanLecPhys2}. For in-plane magnetisation the charges
are located at the film edges. If the film thickness $t$ is much
smaller than the lateral dimensions $L$ and $W$, we can approximate
the charge densities by line charges $\lambda$=$M_\text{s}t$ [A]. The
energy of such a line charge in the magnetic field potential of the
opposite line charge is

\begin{equation}
\label{eq:U}
U=\frac{1}{2}\mu_0\int\lambda\phi dl,
\end{equation}

where the integral is taken along the line. For a magnetisation along the $x$ direction, as shown in
figure~\ref{fig:inplane}, the magnetic potential can be calculated by
a summation of the potential of individual point charges:

\begin{equation}
  \phi_\pm(x,y)=\pm\frac{\lambda}{4\pi}\int_{-\nicefrac{L}{2}}^{\nicefrac{L}{2}}\frac{1}{r_\pm(y')}dy'
\end{equation}

where the line charges are located at $x=\pm\frac{W}{2}$. Rewriting
$r$ in terms of $(x,y)$ and solving the integral we obtain

\begin{eqnarray}
\phi_{\pm}(x,y)=\nonumber\\
\pm\frac{\mu_0}{4\pi}\ln
\left(
 \frac{(y-\nicefrac{L}{2})\sqrt{(y-\nicefrac{L}{2})^{2}+(x+\nicefrac{W}{2})^{2}}}
        {(y+\nicefrac{L}{2})\sqrt{(y+\nicefrac{L}{2})^{2}+(x-\nicefrac{W}{2})^{2}}}
\right)
\end{eqnarray}

Taking the reference potential at (0,0) for both $\phi_+$ and
$\phi_-$, and solving integral~(\ref{eq:U}), we obtain for the magnetic
energy for a field along $x$

\begin{equation}
  U_x=\frac{\lambda^{2}\mu_0}{2\pi}\left(W-\sqrt{L^2+W^2} \right)
\end{equation}

The energy for a field applied along $y$, $U_y$, can simply be obtained by
reversing $W$ and $L$, and we obtain for the energy difference $U_x-U_y$, and
therefore the maximum in-plane torque,

\begin{equation}
\Gamma_{\Vert\text{max}}=\frac{\lambda^2\mu_0}{2\pi}\left(W-L\right)
\end{equation}

As expected, the maximum torque is zero for a square element. 

After replacing $\lambda$ with $M_st$, we can compare the in-plane and
perpendicular torque: 

\begin{equation}
\Gamma_{\Vert\text{max}}=\Gamma_{\perp\text{max}} \left(\frac{W-L}{W}\right)\frac{t}{L\pi}
\end{equation}

Since $t\ll L$ (and $W$), the in-plane torque is much smaller than the
perpendicular torque. For the elements we designed
(table~\ref{tab:DesignRectangular}), the maximum ratio is at most
\num{2.4e-3}, so the out of plane demagnetisation factor is to a very
good approximation equal to unity. Therefore the in-plane difference
in demagnetisation factors can be determined from

\begin{equation}
\Delta N=\left(\frac{W-L}{W}\right)\frac{t}{L\pi},
\end{equation}

and the field dependent torque becomes

\begin{align}
  \label{eq:fielddependence}
  \Gamma&=\Gamma_{\Vert\text{max}}2h\sqrt{1-h^2}\\
  h&=\frac{H}{\Delta N M_\text{s}}.
\end{align}

For the \SI{2.5}{mT} field applied in our experiments, $h<\nicefrac{1}{\sqrt{2}}$ for all
elements and the maximum torque is not attained.

\subsection{Magnetic torque of magnetotactic bacteria}

In contrast to the microfabricated elements, the magnetosome of
magnetotactic bacteria is less well defined. As a first approximation,
we can consider the chain of magnetic spheres as a chain of $n$
dipoles separated at distance $a$, each with a dipole moment
$m$=$M_sV$ [Am$^2$], where $V$ is the volume of each single
sphere. Again, we assume that all dipoles are aligned parallel to the
field ($\varphi=\theta$) to obtain an upper limit on the torque. (See
figure~\ref{fig:AxisMandH} for angle definition). The magnetic energy
for such a dipole chain has been derived by Jacobs and
Bean~\cite{Jacobs1955}, which rewritten in SI units is

\begin{align}
  U=&\frac{\mu_0m^2}{4\pi a^3} n K_n\left(1-3\cos^2(\theta)\right) + \nonumber\\
  &\mu_0nmHcos(\varphi-\theta)\\
 K_n=&\sum_{j=1}^n\frac{(n-j)}{nj^3}
\end{align}

The maximum torque equals the energy difference between the state
where all moments are parallel to the chain ($\theta$=0) and the state
where they are perpendicular to the chain ($\theta$=$\pi/2$):

\begin{align}
  \label{eq:22}
  \Gamma_\text{max}&=\frac{1}{2}\mu_0M_\text{s}^2nV \Delta N\\
  \label{eq:23}
  \Delta N=N_\perp-N_\Vert&=\frac{1}{4}K_n.
\end{align}

As expected, for a single dipole, $n=1$ and there is no energy
difference. The expression for the field dependence of the torque is
equal to that of the microfabricated elements
(equation~\ref{eq:fielddependence}).

\begin{figure}
  \begin{centering}
    \includegraphics[width=\widefigurewidth]{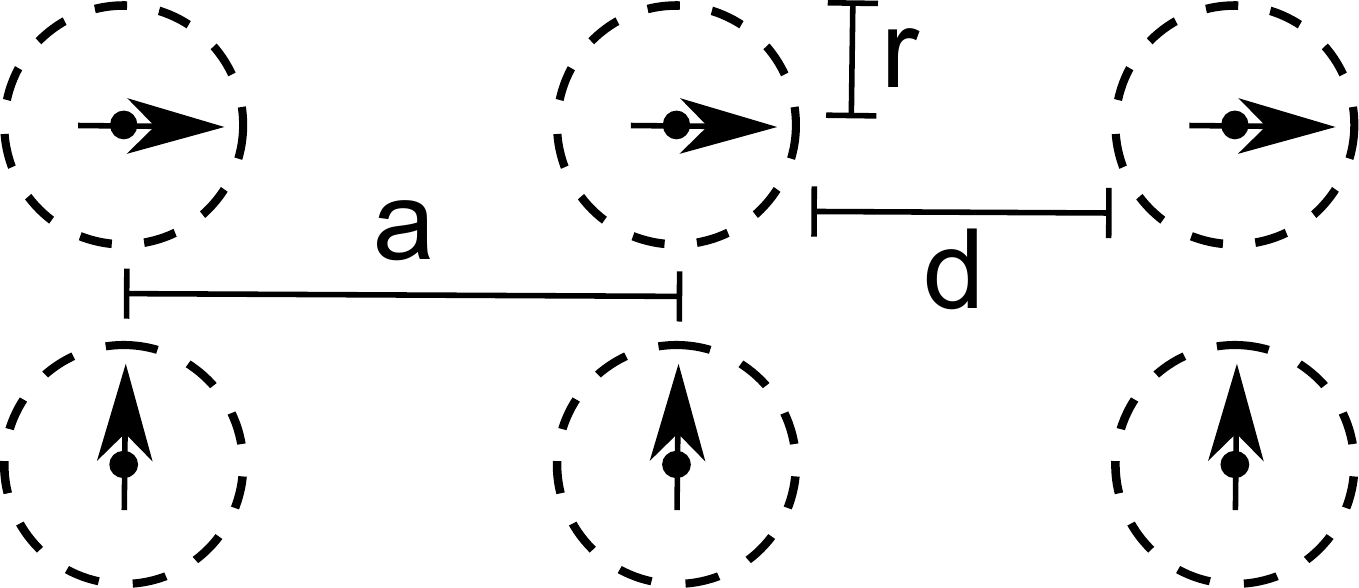}
  \end{centering}
  \caption{Chain of magnetic spheres of radius $r$, spaced by distance
    $d$, approximated
    by point dipoles spaced by a distance $a=r+d$, magnetised along the
    longitudinal axis of the chain (top) or perpendicular to its
    longitudinal axis (bottom).}
  \label{fig:dipoles}
\end{figure}

The magnetosome does not consist of point dipoles but should be
approximated by spheres with
radius $r$, spaced at distance $d$ from each other
(figure~\ref{fig:dipoles}). Naively, we could simply modify the Jacob
and Bean model by setting

\begin{equation}
  a=\frac{d}{r}+2,
\end{equation}

leading to

\begin{eqnarray}
\label{eq:25}
    \Delta N=\frac{2K_n}{\left(\frac{d}{r}+2\right)^3}
\end{eqnarray}

as a correction of equation~\ref{eq:23}.  It is not immediately clear
that this naive correction is valid. Intuitively, one could assume
that the field of a uniformly magnetised sphere is identical to a
dipole field. We ignore however that the field of neighbouring dipoles
is not uniform over the volume of the sphere. To investigate this
effect, we performed finite element simulations.

\subsubsection{FEM calculations}

\begin{figure}
\begin{centering}
\includegraphics[width=\figurewidth]{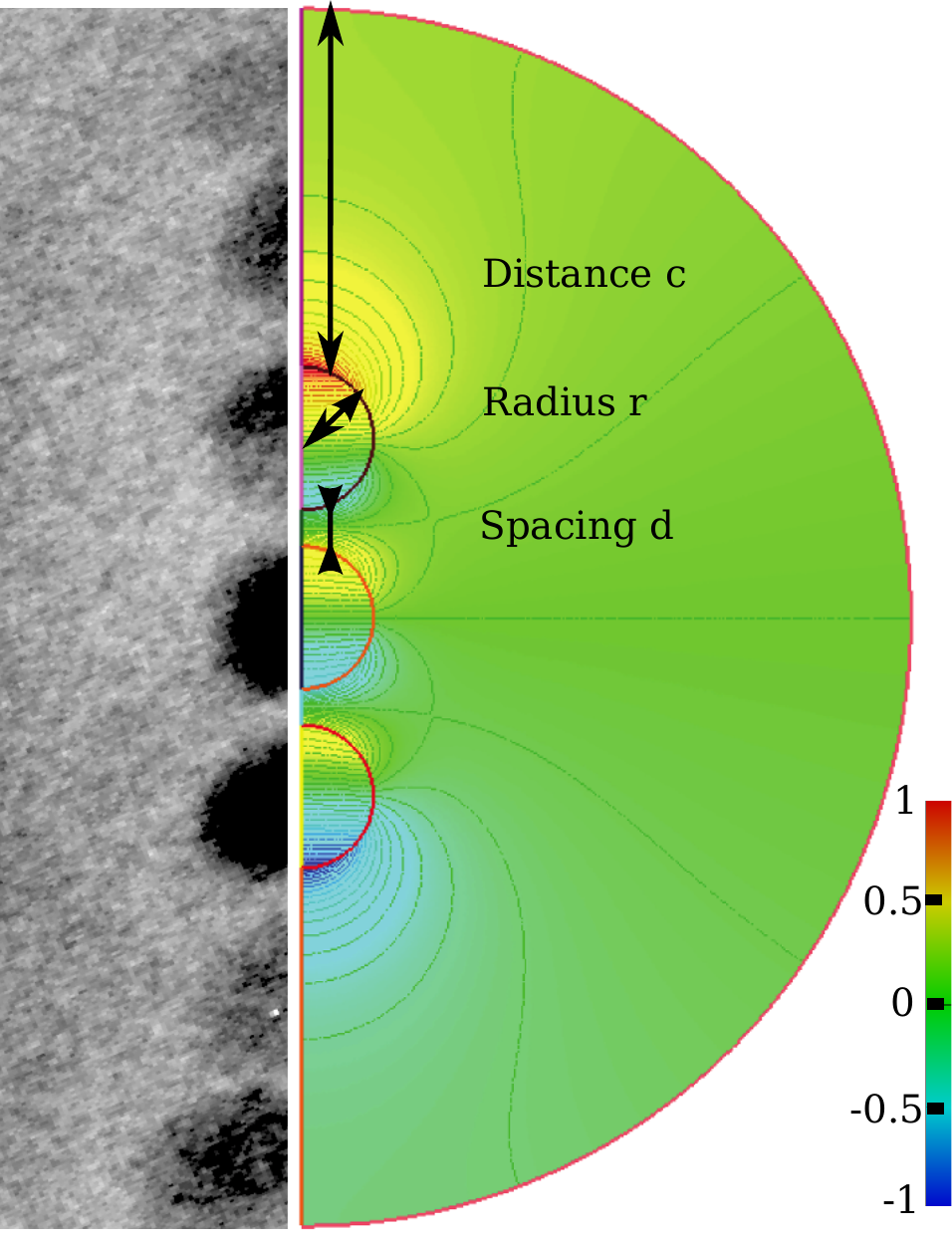}
\end{centering}
\caption{FEM results of $n$ magnetic spheres spaced distance $d$ apart
  at distance $c$ from the edge of the simulation area. In this case
  $n$=3, $r$=1, $d/r$=0.2 and $c/r$=5 (chosen for convenient
  display). The color code and contour lines indicate the magnetic
  scalar potential [A] using a sphere magnetisation of
  \SI{1}{A/m}. On the left, part of the TEM image of figure~\ref{fig:TEM}
  illustrates the magnetosome.}
\label{fig:n_spheres}
\end{figure}

Since there are no free currents, $\rot{H}=0$ and the magnetic field can be
expressed by a scalar magnetic potential field $\bs{H}=-\grad{\phi}$
[A/m] and accompanying magnetic charge $\diver{H}=\rho=-\diver{M}$
[A/m$^2$]. The partial differential equation to be solved in a
cylindrical coordinate system than becomes

\begin{equation}
x\frac{\partial^{2}\phi(x,y)}{\partial x^{2}}+x\frac{\partial^{2}\phi(x,y)}{\partial y^{2}}=\rho(x,y).
\end{equation}

where $x$ is the radial coordinate and $y$ is along the axis. This differential equation was solved in FreeFEM++~\cite{FreeFEM}, see
for example figure~\ref{fig:n_spheres}. We assume a uniform
magnetisation along the axis of the chain of spheres, which leads to a
sinusoidal magnetic surface charge distribution on each semi circular
boundary ($\sigma=\textbf{M}\cdot\textbf{n}$ [A/m]). Integrating
$\boldsymbol{m}\times\boldsymbol{H}$ over the surface of each semi
circle times $2\pi$ (as a cylindrical coordinate system is used), the
demagnetisation energy of the magnetic chain is acquired. By dividing
the demagnetisation energy by the summed volume of all spheres in the
chain the demagnetisation factor for fields applied along the axis of
the chain ($N_\parallel$) is found. Since the system is cylindrically
symmetric and the sum of the demagnetisation factors equals unity,
$N_\parallel+2N_\perp=1$, so

\begin{equation}
  \Delta N= \frac{1}{2}\left(1-3N_\parallel\right).
\end{equation}

To obtain generic results, the dimensions were scaled to $r$ and
potential to $\rho$. Numerical calculations were performed with varying
mesh densities of 6, 10, 14 and 18 points per unit length. The final
value for the demagnetisation factor is obtained by extrapolating to
infinite mesh densities. The distance \emph{C} from the first and last
sphere of the chain to the boundary is varied and chosen such that no
significant effect of the boundary on the simulation results is
observed. Under these conditions, the numerical calculation was found
to be accurate within 0.1\% for the demagnetisation factor of a single
sphere ($\frac{1}{3}$) .

\begin{figure}
\begin{centering}
\includegraphics[width=\widefigurewidth]{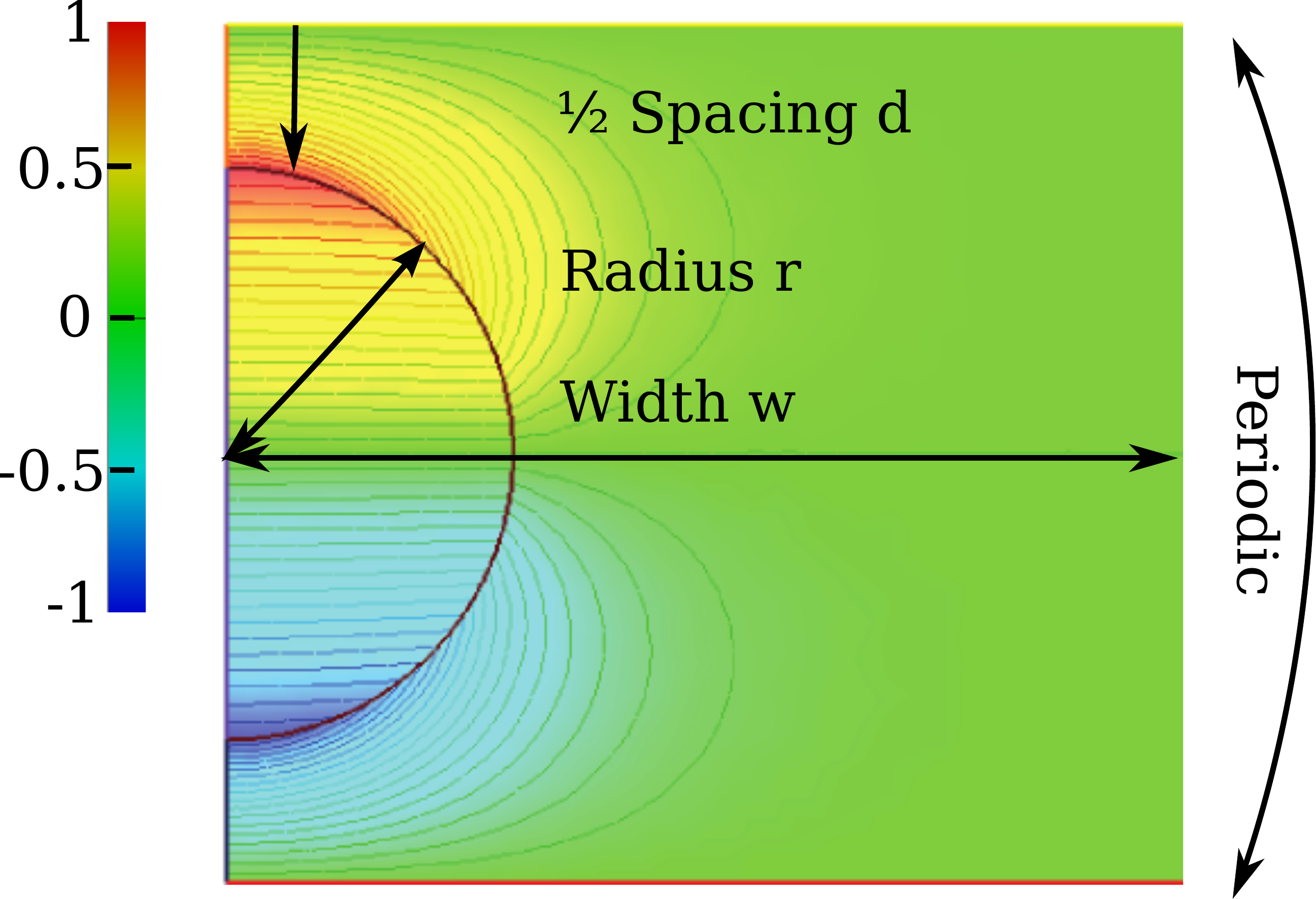}
\end{centering}
\caption{FEM results of an infinite chain of spheres spaced distance
  $d$ apart.  The top and bottom side of the cylindrically symmetric,
  rectangular simulation domain are periodic boundaries. In this case
  $d/r$=1 (of which only half falls in the simulation space),
  $w/r$=2.5 (chosen small for convenient display).}
\label{fig:PeriodicBoundary}
\end{figure}

To find the demagnetisation energy of an infinite chain of spheres,
one semi circle in a rectangular geometry is defined, see figure~\ref{fig:PeriodicBoundary}, with periodic boundary conditions on
the top and bottom boundaries of the rectangle. This creates an
infinite chain of spheres with radius \emph{r}=1, spaced \emph{d}
apart. The periodic boundary ensures that the field, which is leaving
the simulation space from the top, enters the simulation space from
the bottom, and vice versa. The demagnetisation energy is found in a
similar manner as for the FEM calculation of \emph{n} spheres spaced
\emph{d} apart.

\begin{figure}
  \begin{centering}
    \includegraphics[width=\widefigurewidth]{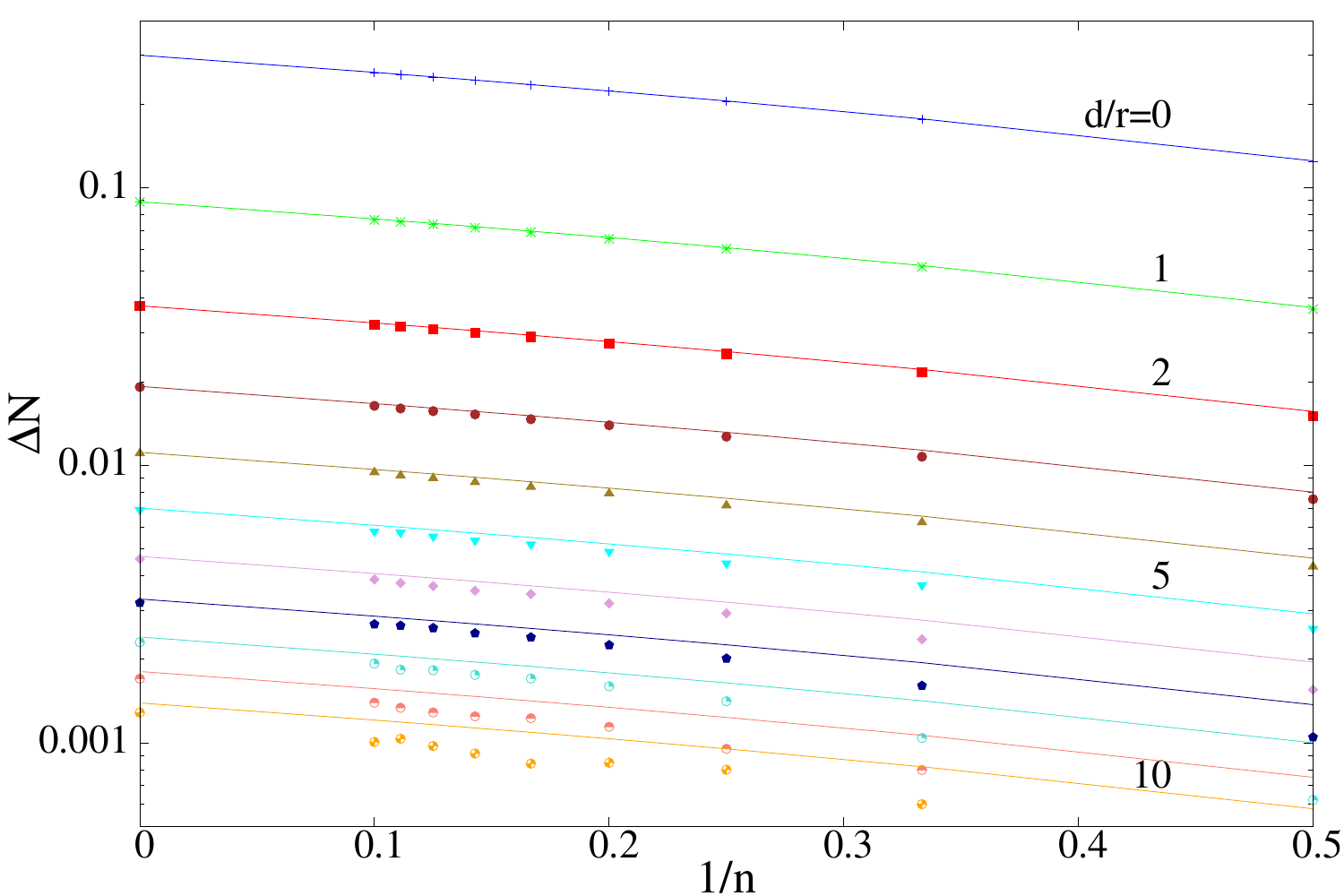}
  \end{centering}
  \caption{Difference in demagnetisation factors of a chain of spheres
    as function of number of spheres $n$ for varying spacing between
    the spheres $d/r$. Shown are FEM calculations (symbols) and the
    Jacobs and Bean model (lines).\cmtr{The color yellow will be
      changed once my gnuplot is fixed :(}}
  \label{fig:Comparison_cal}
\end{figure}

The value of $\Delta N$ obtained from the numerical
calculations is shown in figure~\ref{fig:Comparison_cal}, together
with the naive correction to the Jacobson and Bean chain of spheres
model. The number of spheres $n$ was varied from 2 to $\infty$, and
sphere spacing $d/r$ was varied from 0 to 10. For demagnetisation
factors $>0.01$ the chain of spheres model agrees within 3\% with the
outcome of numerical calculation.  The  increase in deviation
at smaller demagnetisation factors is due to numerical errors in the
FEM calculation.

We therefore conclude that the chain of spheres model provides a very
good, if not exact, solution. It is therefore safe to use
equation~\ref{eq:25} in the calculation of the magnetic torque (eq
\ref{eq:22}) on a magnetism. 

For an infinitely long chain of touching spheres, $d$=0 and
$n\rightarrow\infty$, the difference in demagnetisation factors
($N_\perp-N_\Vert$) approaches $0.3$. Approximating the chain with an
long cylinder ($N_\perp-N_\Vert$=0.5)~\cite{Hanzlik1996, Erglis2007},
overestimates the maximum torque by 40\%. Simply taking the total
magnetic moment to calculate the torque, as if $N_\perp-N_\Vert$=1,
would overestimate by a factor of three.

The FEM calculation shows a property of magnetostatic fields of
spherical objects that was unknown to the authors. Apparently
interaction energy between two uniformly magnetised spheres can be
simplified by mere dipole interaction, leading to the collapse of a six-fold integral
into a simple product. This is very similar to the center of mass concept in
gravitational and electrostatic interaction.

\subsection{Rotational Drag torque}

The torque exerted by the magnetic field results in the rotation of
the microfabricated elements or magnetotactic bacteria. This rotation
on its turn causes a counter torque, which increases with increasing
rotational velocity. An equilibrium steady state velocity is obtained
when the magnetic torque is balanced by the drag torque. Since we work
with very low mass objects, this steady state is reached almost
immediately in comparison to the duration of the experiment. In the
following we will discuss the drag torque for magnetotactic bacteria
and microfabricated elements, as well as the resulting rotational velocity.

\subsubsection{Magnetotactic bacteria}
magnetotactic bacteria are very small, and rotate at a few
revolutions per second only. Inertial forces therefore do not play a
significant role. The ratio between viscous and inertial forces is
characterised by the Reynolds number $R$, which for rotation at angular
velocity $\omega$ [rad/s] is

\begin{equation}
R=\frac{L^2\rho\omega}{4\eta},
\end{equation}

where $L$ is the characteristic length (in our case the length of the
bacterium), $\rho$ the density and $\eta$ dynamic viscosity
of the liquid (for water respectively
\SI{e3}{kg\per \cubic \meter}, and \SI{1}{mPas}). Experiments by Dennis
\emph{et al}~\cite{Dennis1980} show that a Stokes flow approximation for
the drag torque is accurate up to $R$=$10$. In experiments with
bacteria, the Reynolds number is the order of \num{e-3} and Stokes
flow approximation is certainly allowed. The drag torque is therefore

\begin{equation}
\label{eq:rotational drag}
\Gamma_{\textrm{D}}=f_{\textrm{r}}\omega,
\end{equation}

where the drag coefficient $f_\text{r}$ can be approximated by

\begin{equation}
f_{\textrm{r}}=\frac{\pi\eta L^{3}}{3\ln\left(\frac{2L}{W}\right)-\frac{3}{2}}
\end{equation}

assuming that the bacterium is a prolate spheroid with length $L$ and diameter
$W$~\cite{Berg1993}. 

\subsubsection{Microfabricated elements}
In principle, one could fabricate elements using lithographic
techniques of identical size to magnetotactic bacteria. The handling
of such elements will be however be cumbersome, since one cannot use
tweezers at that scale. When scaling up however, we should remain in
the Stokes flow regime for the viscous drag model to remain valid. Therefore we increased
the size of the elements up to the point that their Reynolds number equals
approximately unity, well below $10$ so that Stokes flow approximations
can still be assumed. 

We approximate the rectangular elements with a disc of diameter equal
to the length of the element $L$~\cite{Berg1993}.  Since the elements are floating on
the water surface, we further assume that we can simply divide the drag
coefficient by two, since drag will only occur on one side of the
element. Under these assumptions, the drag coefficient is

\begin{equation}
f_{\textrm{r}}=\frac{1}{2}\left[ \frac{32}{3} \eta \left(\frac{L}{2}\right)^3\right]=\frac{2}{3}\eta L^{3}.
\end{equation}

The assumptions are rather crude, but not significant compared to
measurement errors.

\subsection{Maximum rotation frequency}

Finally, the maximum steady-state rotational speed will be obtained when
the maximum magnetic torque equals the rotational drag torque

\begin{equation}
\label{eq:maxrotfreq}
\omega_\text{max}=\frac{\Gamma}{f_\text{r}}
\end{equation}

\section{Experimental}

\subsection{Microfabrication}
\begin{figure}
  \begin{centering}
    \includegraphics[width=\figurewidth]{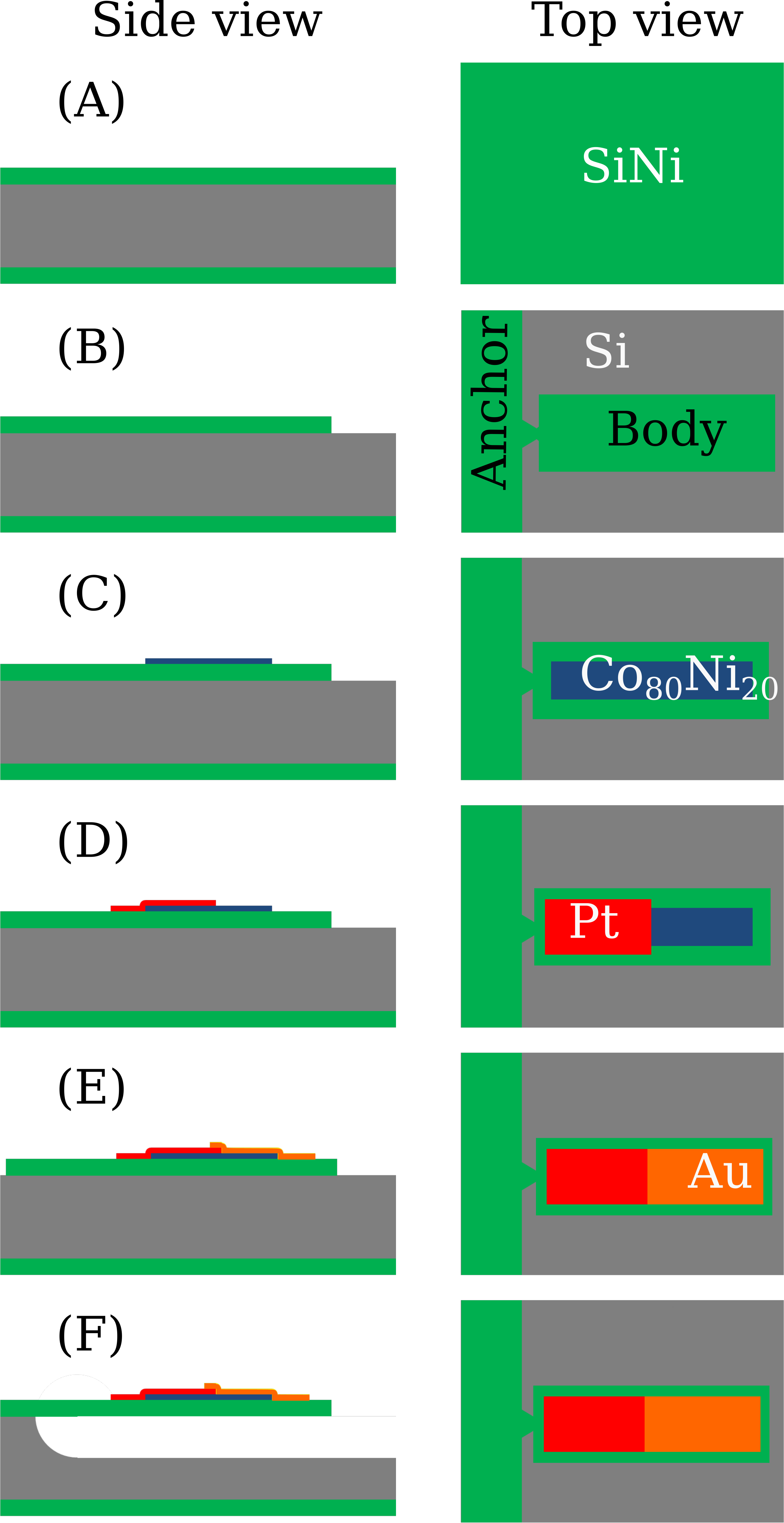}
  \end{centering}
  \caption{Side view (left) and top view (right) of the process flow used to
    fabricate silicon-nitride elements with a magnetic strip. See text
  for a description of the process steps.}
  \label{fig:fabrication}
\end{figure}

The microfabricated elements are realised by surface micro machining techniques on a
100 mm diameter p-type silicon wafer (orientation $<$100$>$). The fabrication steps
are illustrated in figure \ref{fig:fabrication}. A four mask
process is used with masks numbered I to IV. 

\begin{figure}
  \begin{centering}
    \includegraphics[width=\figurewidth]{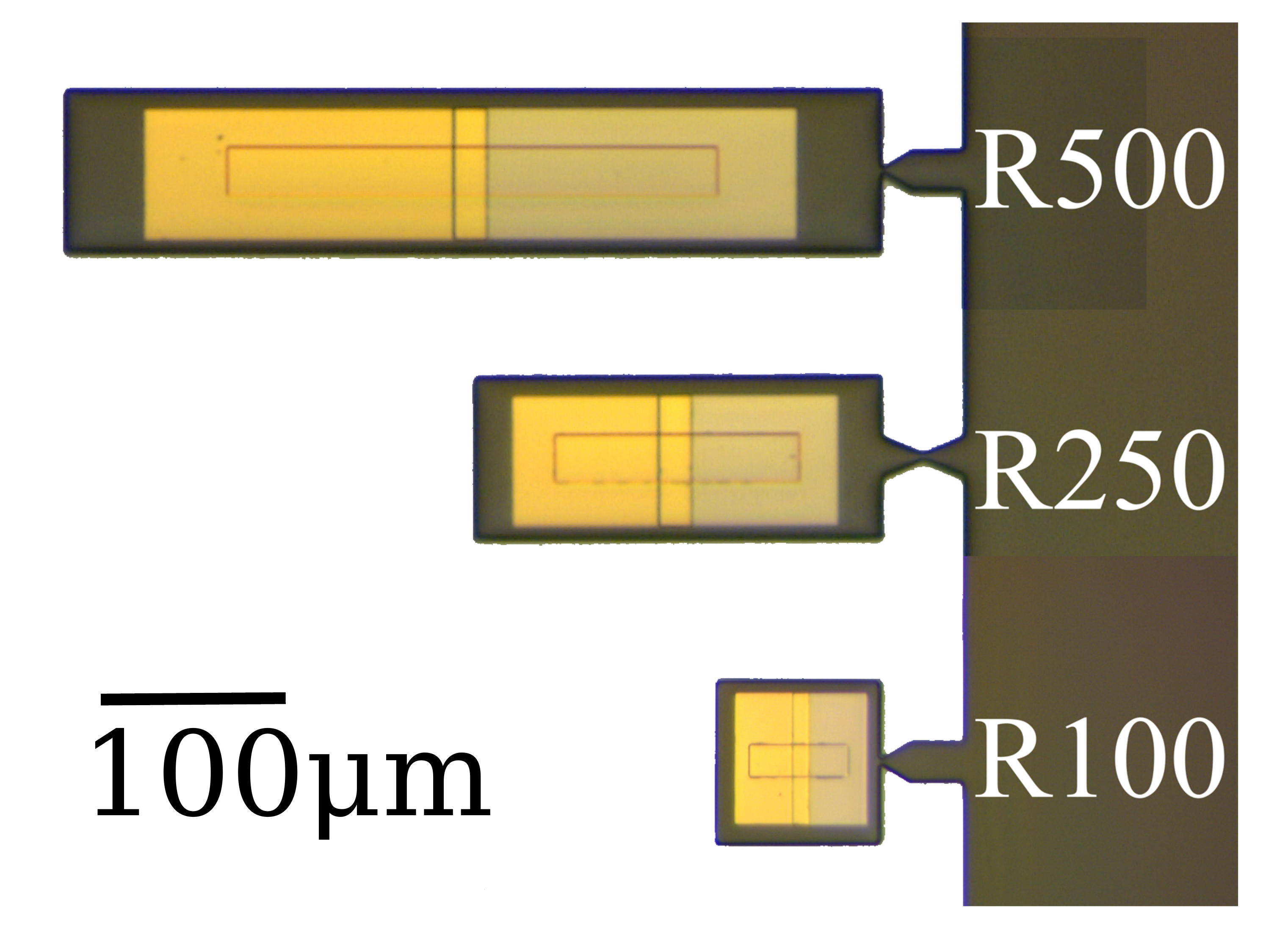}
  \end{centering}
  \caption{Three rectangular elements of different length (500,
250 and \SI{100 x 100}{\micro m}), still attached to their
support. These elements are broken off by a tweezer and placed on the
water surface.} 
\label{fig:MFS}
\end{figure}

Fabrication starts with the deposition of 2-3 \si{\um} low-stress
nitride in a LPCVD process (step (A)). Next, a photoresist mask is
patterned using Mask I and the unprotected nitride is removed with RIE
using a plasma of CHF$_3$ and O$_2$. The excess resist is removed in
a oxygen plasma (B). A new layer of photoresist is patterned using Mask II
after which a \SI{200}{nm} thick Co$_{80}$Ni$_{20}$ layer is deposited
using e-beam evaporation to define the magnetic strips. A lift-off
process (C) is used to remove the excess metal and photoresist. Using
Mask III a new layer of photoresist is patterned and after sputtering
\SI{75}{nm} Pt a similar lift-off process is used (D). This lift-off
procedure is repeated using Mask IV after sputtering a \SI{75}{nm} Au
layer (E). The Pt/Au structures were intended for experiments with
self-propelled structures in peroxide solution~\cite{Ismagilov2002}, which failed however
and are not reported in this paper. Using Mask I again a protective layer of
photoresist is patterned to protect the nitride/metal structures
during XeF$_2$ etching of silicon (at \SI{530}{Pa} pressure). Finally
the photoresist is removed using O$_2$ plasma etching (F). The free
hanging nitride structures with metal layers on top are still anchored
to the substrate but can be easily snapped off using tweezers or a
needle (Figure~\ref{fig:MFS}). 

Table~\ref{tab:DesignRectangular} lists the dimensions of the nitride
supports ($W,L$) and their magnetic strips ($w,l$) for the elements
used in the rotation experiments. Table~\ref{tab:TMM} list the array
of elements that was used for the torque magnetometry
experiments.

\begin{table}
    \caption{Microelements design (width and length  of
      silicon-nitride elements ($W$, $L$) and magnetic elements ($w$,
      $l$)), drag coefficient $f_\text{r}$, maximum torque
      $\Gamma_{\parallel\text{max}}$, effective field $h$ and torque
      $\Gamma_{\parallel}$ for a field of \SI{2.5}{mT}} 
    \label{tab:DesignRectangular}
    \begin{ruledtabular}
    \begin{tabular}{cccccccc}
      $W$ & $L$  & $w$ & $l$ & $f_\text{r}$ & $\Gamma_{\parallel\text{max}}$ &
      $h$ &$\Gamma_{\parallel}$\\
       {[}\si{\micro m}] &  [\si{\micro m}] & [\si{\micro m}] &
       [\si{\micro m}]  & [fNms] & [pNm] &  &[pNm]\\
      25 & 100 & 5 & 60 & 0.7 & 0.6 & 0.14 &0.09\\
      25 & 250 & 5 & 150 & 10 & 1.6 & 0.14 & 0.22\\
      25 & 500 & 5 & 300 & 83 & 3.3 & 0.13 & 0.44\\
      50 & 100 & 15 & 60 & 0.7 & 0.5 & 0.52&0.23\\
      50 & 250 & 15 & 150 & 10 & 1.5 & 0.44&0.60\\
      50 & 500 & 15 & 300 & 83 & 3.2 & 0.41&1.21\\
    \end{tabular}
    \end{ruledtabular}
\end{table}

\begin{table}
    \caption{Elements in array measured by TMM}
    \label{tab:TMM}
    \begin{ruledtabular}
    \begin{tabular}{ccc}
      \toprule 
      Amount & Area nitride [\si{\micro m}] & Area \CoNi  [\si{\micro m}]\\
      \midrule 
      68 & 500 $\times$ 100 & 300 $\times$ 30\\
      68 & 250 $\times$ 100 & 150 $\times$ 30\\
      34 & 100 $\times$ 100 & 60 $\times$ 30 \\
      \bottomrule
    \end{tabular}
    \end{ruledtabular}
\end{table}

\subsection{Growth of magnetotactic bacteria}
The magnetotactic bacteria used were \emph{Magnetospirillum
  magnetotacticum}, MS-1, isolated by ATCC (31632), purchased through
LGstandards and delivered on dry ice. The growth medium is prepared
using the instructions from ATCC in accordance
with~\cite{Paoletti1988} and~\cite{Blakemore1979}.  The frozen culture
is placed in the growth medium and left to defrost in an
\SI{16x125}{mm} vial, kept at \SI{27}{\celsius}. The bacteria are
harvested after an incubation period of two weeks by centrifuging the
sample at \SI{7000}{rpm} and discarding the supernatant liquid. Part
of the harvested MTB are suspended on new growth medium for future
use. 

\begin{figure}
  \begin{centering}
    \includegraphics[width=\figurewidth]{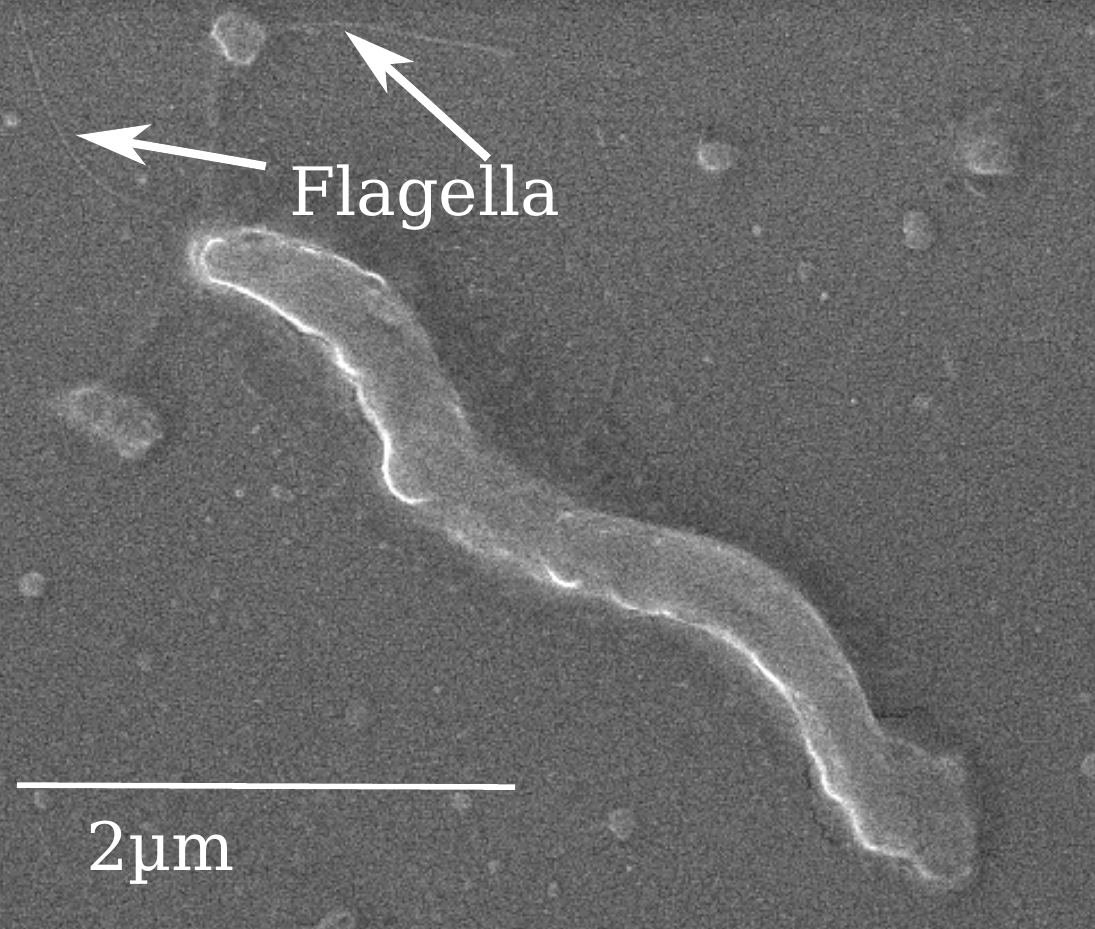}
    \includegraphics[width=\figurewidth]{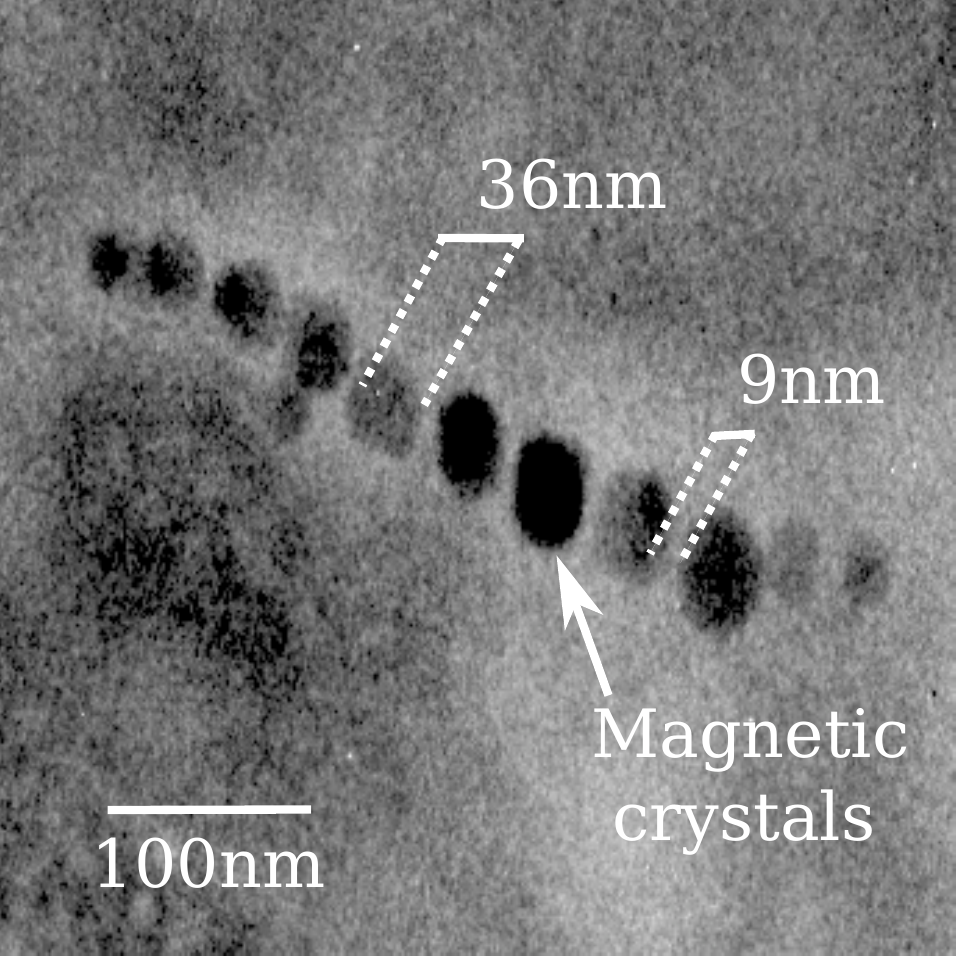}
  \end{centering}
  \caption{Top: SEM image of a magnetotactic bacterium. The flagellae
    can be observed at the top left. Bottom: TEM image of the
    magnetosome chain. }
\label{fig:TEM}
\end{figure}

The magnetotactic bacteria were observed in a Scanning Electron
Microscope (SEM, JEOL 6510) and a Transmission Electron Microscope
(TEM, Gatan / FEI EFTEM). Figure~\ref{fig:TEM} shows a SEM image of a single
magnetotactic bacterium with its flagella, as well as a TEM image of
the magnetosome chain.

\subsection{Torque magnetometry}
After fabrication, the microfabricated elements elements are still
anchored to the substrate and torque magnetometry measurements can be
performed. Since the torque of individual elements is far below the
sensitivity of the magnetometer, measurements were done on an array
(table~\ref{tab:TMM}). But even on an array of over 100 elements the
in-plane torque is too small to be detected. Therefore we only
investigated the torque for the magnetic field rotating out of the
film plane.

Torque magnetometry was performed on a home-built instrument which
measures the torque on a sample of magnetic material as a function of
the magnitude and angle of the external magnetic field. The sample is
attached to a glass rod suspended from a thin torsion wire and
subjected to a uniform magnetic field generated by an electromagnet
that can be rotated around the sample. Rather than measuring the
rotation of the glas rod, the torque applied to the sample by the
electromagnet is compensated by a counteracting torque. For this a
feedback coil is attached to the glass rod, which is placed in the
field of two small permanent magnets. The rotation of the glass rod is
measured by a light beam reflected on a mirror on the glass rod
towards two side-by-side placed photo diodes. The current through the
feedback coil is adjusted such as to keep the focused light beam
exactly between the photo diodes.  This current is logged and, once
calibrated, is an accurate measure of the magnetic torque exerted on
the sample. In this way, any variation in the torsional spring
constant of the suspending wire are eliminated.

\subsection{Optical microscopy}
Experiments with the microfabricated elements and bacteria were
performed in a microscope equipped with a video acquisition system and
a set of coils to generate magnetic fields. Opposite coils are
connected in series, and driven with sinusoidal currents with a
\SI{90}{\degree} phase difference between the orthogonal coil
sets. This generates a uniform, rotating magnetic field.

\begin{figure}
  \begin{centering}
    \includegraphics[width=\widefigurewidth]{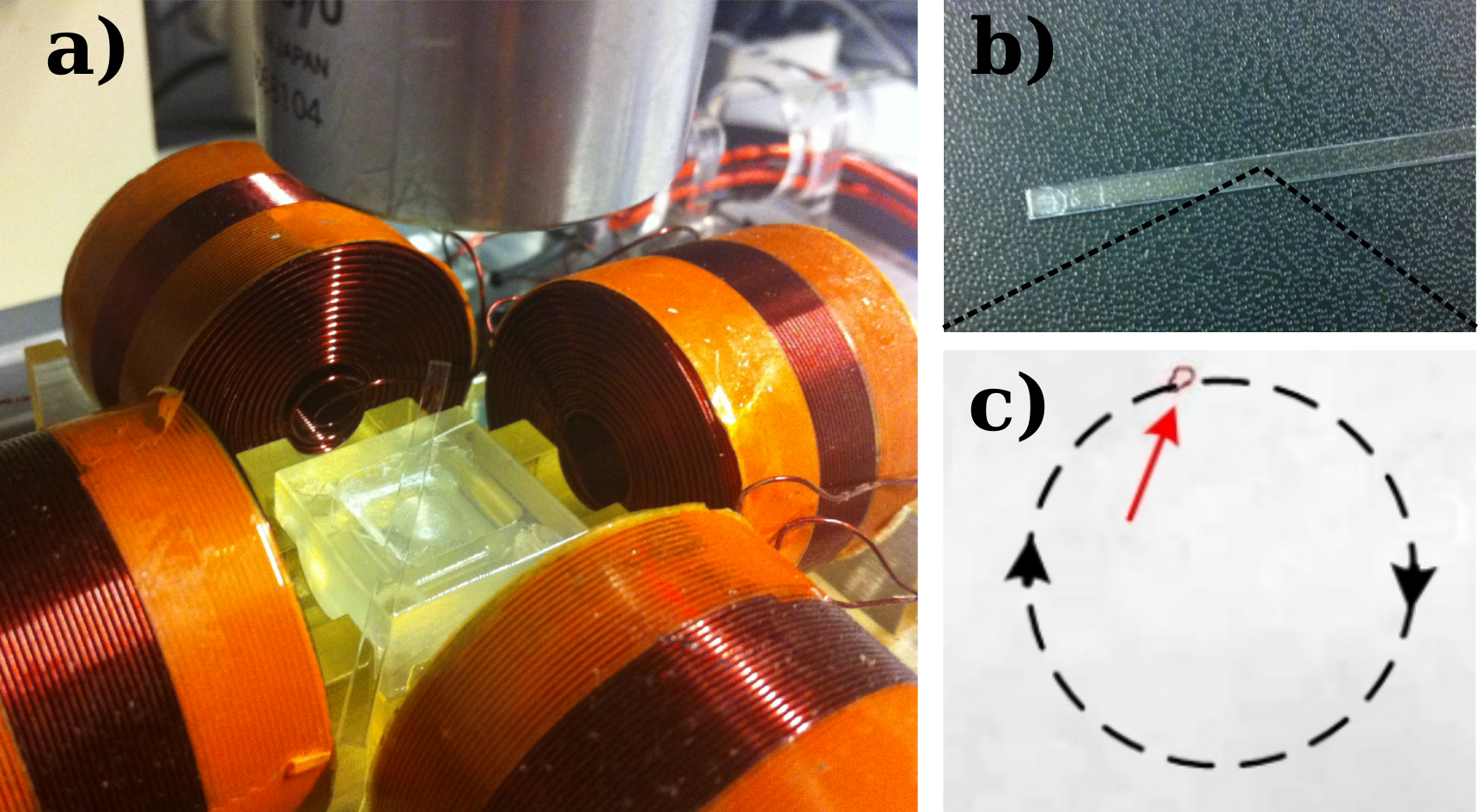}
  \end{centering}
  \caption{Measurement setup. a) Four coils are used to generate an
    in-plane magnetic field of arbitrary direction b) The
    magnetotactic bacteria are inserted in a cuvette c) A microscope
    with video-capture system and tracking software is used to monitor
    the movement of the bacteria.}
\label{fig:setup}
\end{figure}

For the microfabricated elements, a small container was fabricated by
3D printing, which could be positioned under the microscope objective
and between the coil system, see figure \ref{fig:setup}. The
microfabricated elements were broken out of the support wafer with a
tweezer and positioned on the water surface. Forcefully submerging the
elements was possible, in which case the elements sank to the bottom
of the container. In that case however the elements stuck to the
container surface and could not be rotated. A rotating magnetic field
of \SI{2.5}{mT} was applied and its frequency was slowly
increased. The frequency at which the elements stopped following the
field and started wobbling was recorded. The experiment was repeated
five times per element.

Experiments on the bacteria are conducted using borosilicate
capillaries (VitroCom, VitroTubes 3520-050, Germany) with a rectangle
channel of \SI{0.2x2}{mm} and a length of \SI{50}{mm}. The capillaries
are filled with growth medium containing magnetotactic bacteria and
the ends of the capillary are sealed with parafilm. The circular
motion (superposition linear movement due to flagella propulsion and
rotary movement due to magnetic torque) of the bacteria was captured
for varying rotation magnetic field frequencies.

\section{Results and Discussion}

\subsection{Torque measurements on microfabricated elements}

\begin{figure}
  \begin{centering}
    \includegraphics[width=\widefigurewidth]{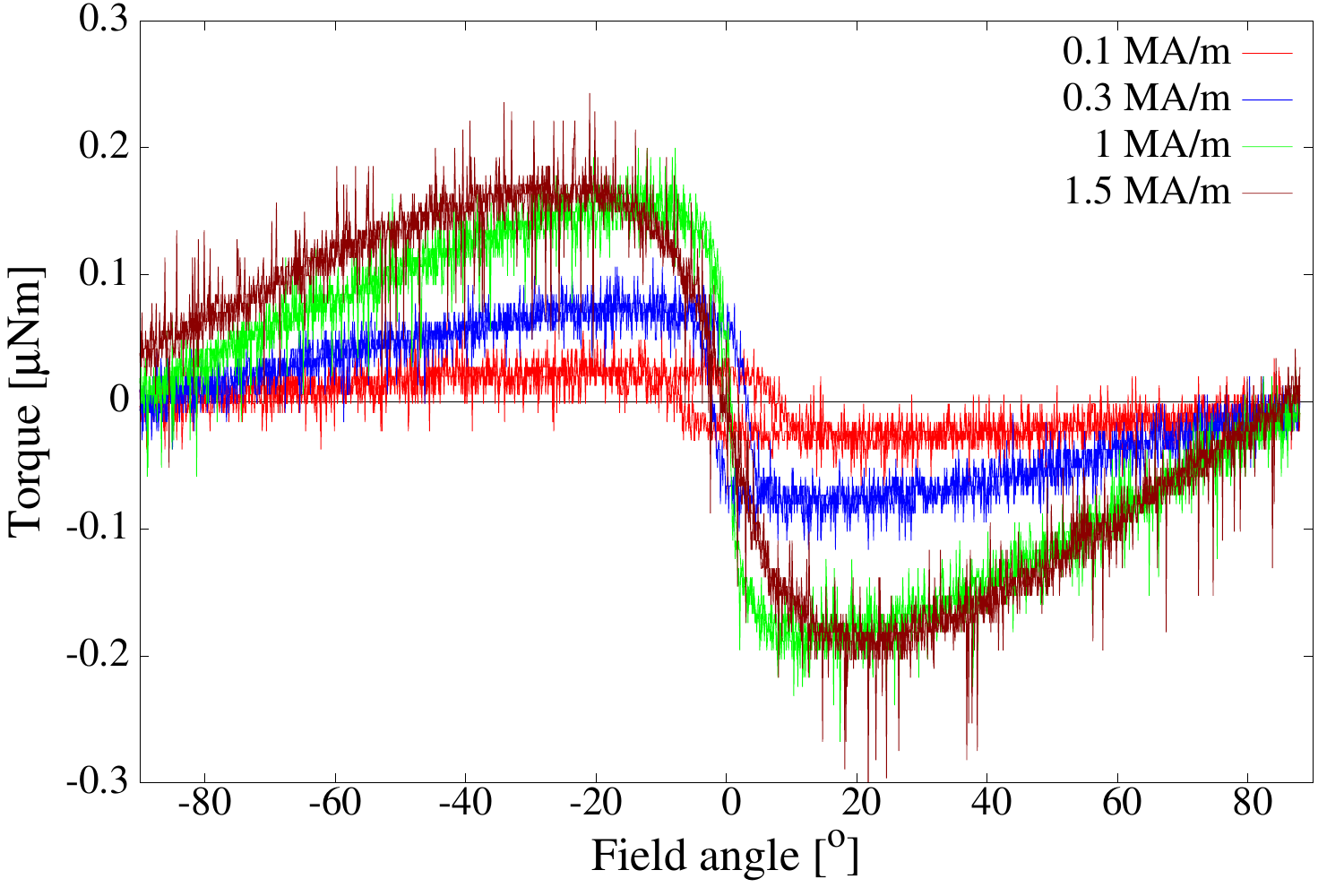}
  \end{centering}
  \caption{Measured out-of-plane torque curves as a function of the field angle
for varying magnetic field strength.}
\label{fig:T_angle}
\end{figure}

Figure~\ref{fig:T_angle} shows the magnetic torque as a function of
applied field angle, where the field is rotating in a plane
perpendicular to the sample surface and parallel to the long axis of
the elements. The signal is at the limit of the instruments
capability, but the torque clearly increases with increasing field
value. At low fields, some hysteresis is visible which indicates that
in zero field the magnetic elements most likely have a remanent
magnetisation.

\begin{figure}
  \begin{centering}
    \includegraphics[width=\widefigurewidth]{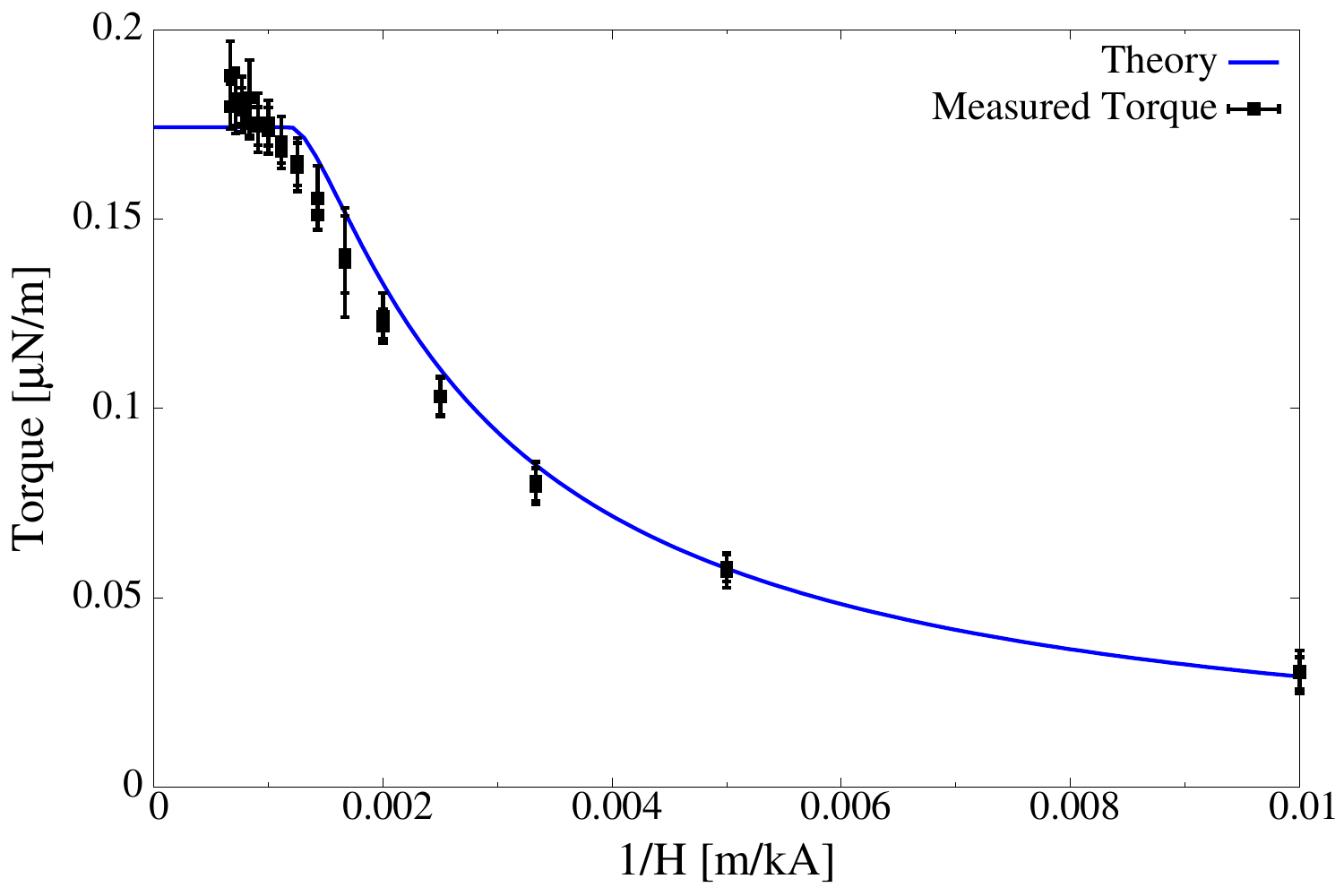}
  \end{centering}
  \caption{Measured maximum out-of-plane torque as a function of
    magnetic field strength compared to the modeled expected behaviour
    (solid line), which does not include any fit parameter.}
\label{fig:T_max_function_H}
\end{figure}

By averaging, an accurate estimate of the maximum torque can be
obtained, which is plotted as a function of the external field in
figure~\ref{fig:T_max_function_H}. Also shown is the theoretical
prediction of the field dependent torque
(section~\ref{sec:perpendicular-fields}), based on the designed lateral
dimensions of the elements, the magnetic layer thickness obtained from
the deposition run and assuming that the composition of the Co-Ni film is
exactly 80/20 which leads to a saturation magnetisation of
\SI{1.19}{MA/m}. No fit parameter has been applied. Considering that
the layer thickness has an estimated error of 5\%, the agreement
between model and experiment is excellent.

\subsection{Maximum rotation frequency of microfabricated elements}

\begin{figure}
  \begin{centering}
    \includegraphics[width=\widefigurewidth]{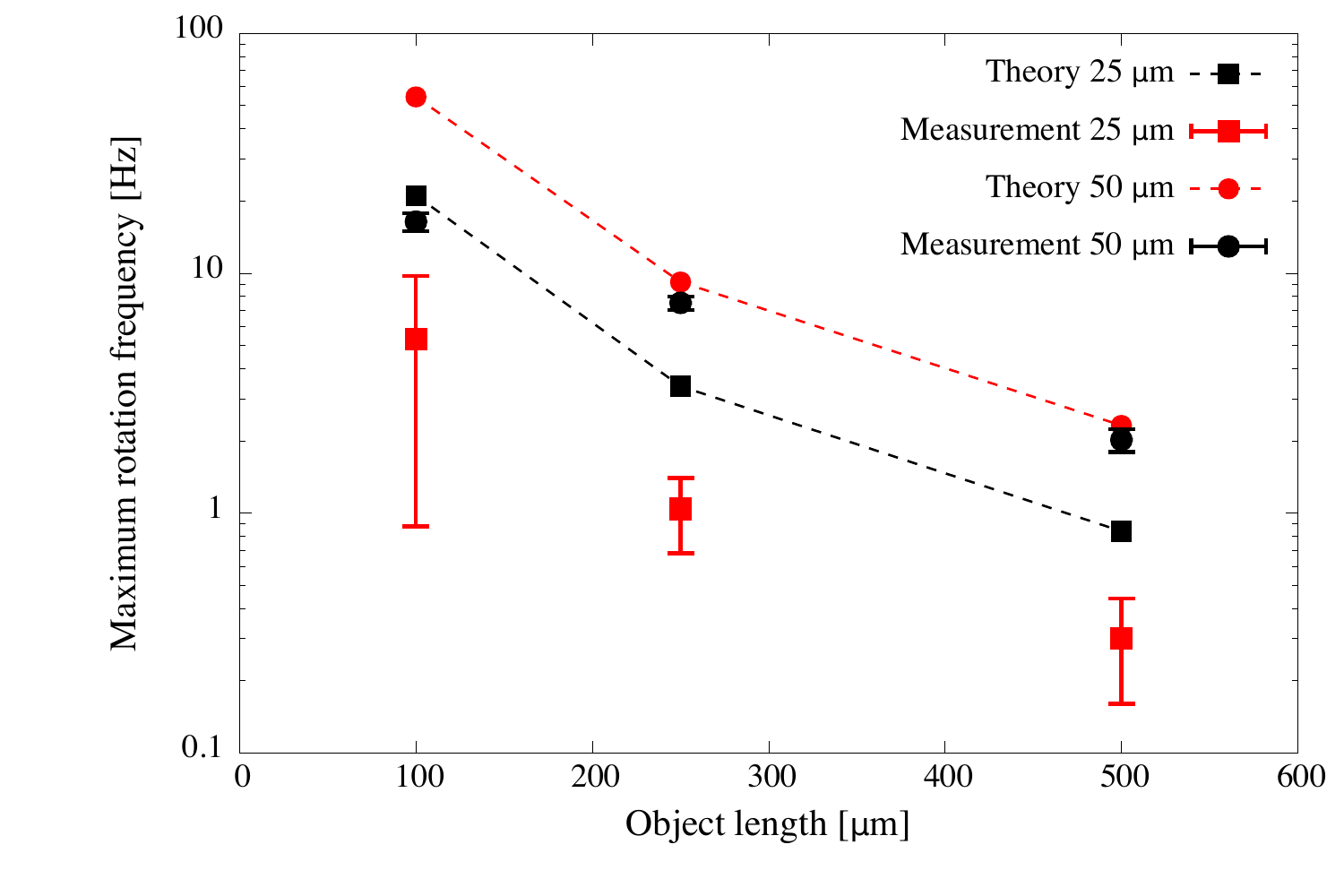}
  \end{centering}
  \caption{Maximum rotation frequency as a function of object length
    for rectangles of 25 (red squares) and \SI{50}{\micro m} (black
    circles) width. (Model points are connected by a line to guide the
    eye, see table~\ref{tab:DesignRectangular}).\cmtr{The
      um will be changed to \si{\micro}{m} as soon as I fix my gnuplot
    :(}}
\label{fig:Fmax_versus_length}
\end{figure}

Figures~\ref{fig:Fmax_versus_length} shows the maximum rotation
frequency for the elements listed in
table~\ref{tab:DesignRectangular}, sorted by the length of the
object. Also shown is the theoretical prediction, using drag
coefficient and magnetic torque as listed in
table~\ref{tab:DesignRectangular}. Since the dimensions of the
magnetic elements is varying with object length, the theoretical
maximum rotation frequency is only calculated for specific elements,
and connected by a line to guide the eye.

For all measurements, the observed maximum rotation frequency lies
below the theoretical prediction. This can only partly be explained by
the measurement procedure, in which the maximum frequency is
approached from below. The difference between the 25 and \SI{50}{\um}
elements is however pronounced. The \SI{50}{\um} elements show small
scatter between experiments, and for the larger element the difference
with theory is small. The \SI{25}{\um} elements however show much more
scatter, and differ from theory by a approximately a factor of
three. 

The difference between the two series is probably not caused by a
difference in drag, which is dominated by the length of the
element. The \SI{25}{\um} elements however have magnetic strips with a
much smaller width, leading to higher saturation fields and relatively
low effective fields ($h$ in table~\ref{tab:DesignRectangular}). We
speculate that the assumption of uniform magnetisation might not hold
for fields that only 15\% of saturation field. Comparing with fields
applied out of the plane, we indeed observe hysteresis effects for
fields below \SI{300}{kA/m}, which is 25\% of the saturation field. It
is possible that the magnetisation in the strip splits up in domains,
leading to a reduction in torque and increase in scatter due to
hysteresis effects. The electromagnet in our setup was unfortunately
not capable of generating higher rotating fields, so this hypothesis
could not be verified.

\subsection{Maximum torque of magnetotactic bacteria}

\begin{figure}
  \begin{centering}
    \includegraphics[width=\widefigurewidth]{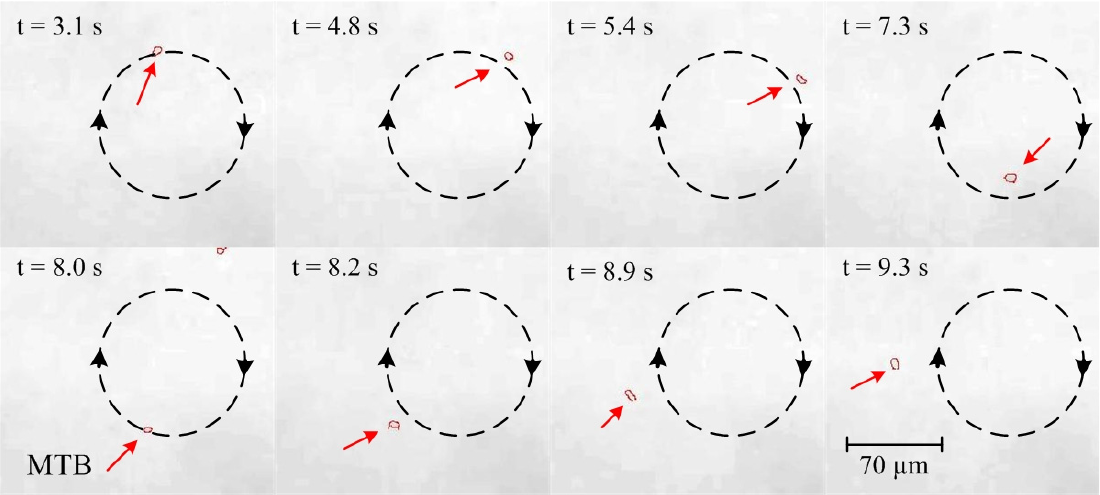}
  \end{centering}
  \caption{Motion of a magnetotactic bacterium under the influence of
    a rotating magnetic field of \SI{7.9}{mT} with gradually
    increasing frequency from 1 to \SI{10}{rad/s}. The black-dashed
    circle illustrates the path and direction of the bacterium (red
    arrow). The images are processed to detect the edges of the
    bacterium. As we gradually increase the rotating field frequency
    to \SI{9.5}{rad/s}, the bacterium can no longer follow the
    field. (Reprinted with permission from~\cite{Khalil2013c})}
\label{fig:RotatingMTB}
\end{figure}

Figure~\ref{fig:RotatingMTB} shows a typical trajectory of a
magnetotactic bacterium under application of a rotating magnetic
field. In the sequence of images, the frequency of rotation is
increased at constant rate until the bacteria no longer follows the
field. In this way we estimate that that maximum rotation frequency
lies in the range fo 7 to \SI{10}{rad/s}. 

From SEM and TEM images of fifteen magnetotactic bacteria
(figure~\ref{fig:TEM}), we determined their dimensions and the size,
spacing and number of crystals in the magnetosomes. The minimum,
maximum and average values are listed in table
\ref{tab:MTB_torque}. Applying the chain-of-spheres model to the
magnetosome data results in $\Delta N$ values of 0.07 to 0.13. As a
result torque values ranging from 0.02 to \SI{4}{aNm}, with an
average of about \SI{1}{aNm} at \SI{7.9}{mT}. 

For the calculation of the field dependent torque, we assume that the
magnetisation of the crystals is that of bulk magnetite (Fe$_3$O$_4$,
\SI{480}{kA/m}~\cite{Witt2005}). Using this value our model predicts that at
\SI{7.9}{mT} the torque has not yet saturated, and could for average
bacteria be increased by a factor five if the field were increased to
$h=\nicefrac{1}{\sqrt{2}}$, so $B$=\SI{59}{mT}. At \SI{7.9}{mT}, the
field is still sufficiently low allow approximation of the torque by
$\bs{m}\times\bs{B}$, with an error of less than 2\%. The error
however increases to at least 50\% at the field at which maximum
torque is reached, and keeps on increasing.  

Unfortunately, the SEM and TEM images could not be performed on
identical bacteria. For the calculation of their theoretical maximum
rotation frequency, we assumed however that the dimensions of the
bacteria and their magnetosomes are correlated. In that case we
predict maximum rotation frequencies to range from 8 to \SI{66}{rad/s}
with an average of 20. Our observation rotation frequency of
\SI{10}{rad/s} lies within that range.

\begin{table*}
  \caption{Minimum, mean and maximum characteristics of 
    magnetotactic bacteria. Length $L$ and width
    $W$ and  amount $n$, radius $r$ and spacing
    $d$ of magnetosomes. From these values the  drag coefficient
    $f_\text{r}$, demagnetisation factors $\Delta N$, magnetic moment $m$ and maximum
    magnetic torque  $\Gamma_{\parallel\text{max}}$ are estimated. For
    a field of \SI{7.9}{mT}, the effective field $h$,  torque $\Gamma_{\parallel}$  and  resulting rotation frequency $\omega$ are calculated. }
  \label{tab:MTB_torque}%
  \begin{ruledtabular}
    \begin{tabular}{lcccccccccccc}
      &  L  & W  & n& r& d 
      & $f_\text{r}$ & $\Delta N$ & $m$ & $\Gamma_{\parallel\text{max}}$  & $h$
      & $\Gamma_{\parallel}$ &$\omega$\\
      & [\si{\micro m}]&  [\si{\micro m}] &  [\#] &  [nm] & [nm] &
      [zNs] &  & [aAm$^2$] & [aNm] &   & [aNm] & [rad/s]\\
      %           L      W       n       r      d      fr      DeltaN m Tmax h  T  w
      min   & 2  & 0.10  &6  & 6   & 6    & 2.6 & 0.07 & 2.61 &0.05 &0.19 & 0.02 & 7.7\\
      mean & 5  & 0.20 &10 & 17 & 9   & 38  & 0.13 & 98.8 &3.8& 0.10   & 0.77 & 20\\
      max   & 6 & 0.25  &16 & 26 & 15 & 67 &  0.13 &566 & 22   & 0.10 & 4.4 & 66\\
    \end{tabular}
  \end{ruledtabular}
\end{table*}

\section{Discussion and outlook}
Using electromagnetic coil systems, it is difficult to achieve high
field values. As a result, our experimental setup did not allow us to
reach the maximum torque that could in principle be applied to both
the microfabricated elements and magnetotactic bacteria. The theory
in section~\ref{sec:theory} provides a simple expression for field
dependent torque, which is verified by the out-of-plane torque
measurements of figure~\ref{fig:T_max_function_H}. This provides a
proven and simple framework to design future experiments.

Our experiments show that too low field values cause irreproducible
results and large deviations from our analytical model. Incorporation
of domain theory into the model will make it far less useable.  It is
therefore advisable to design magnetic field systems that are capable
of higher field values, for instance by using permanent magnet systems
that can mechanically be adjusted to tune field direction and
strength. For magnetotactic bacteria, field values should exceed \SI{60}{mT}.

Correct analysis of the experiments with the microfabricated elements
suffered from the fact that they had to be floated on the water
surface. When submerged however, the elements sink to the bottom,
which again will complicate modelling. Unless one finds a way to
levitate the elements, the influence of interfaces has to be
incorporated in the model. For future experiments, it would be advisable to
carefully take the interface into account. One could for instance
position the elements at the interface between water and oil, which
will provide a more reproducible situation.

The FEM simulations show that the simple, analytical \emph{chain of
  spheres} model accurately predicts the maximum magnetic torque on a
magnetosome. There is no need to approximate the magnetosome with a
solid cylinder, or even worse, a single dipole.

The SEM and TEM observations show that there is a large spread in
dimensions and magnetic properties within a colony of magnetotactic
bacteria. Experimental difficulty prohibits correlation between the
size of a bacterium and it's maximum magnetic torque. In future
experiments, it would be advisable to expand the number of
measurements to determine distributions, or to pre-filter bacteria on
a certain property such as for instance size, speed or minimum radius
of curvature.

\begin{figure}
  \begin{centering}
    \includegraphics[width=\widefigurewidth]{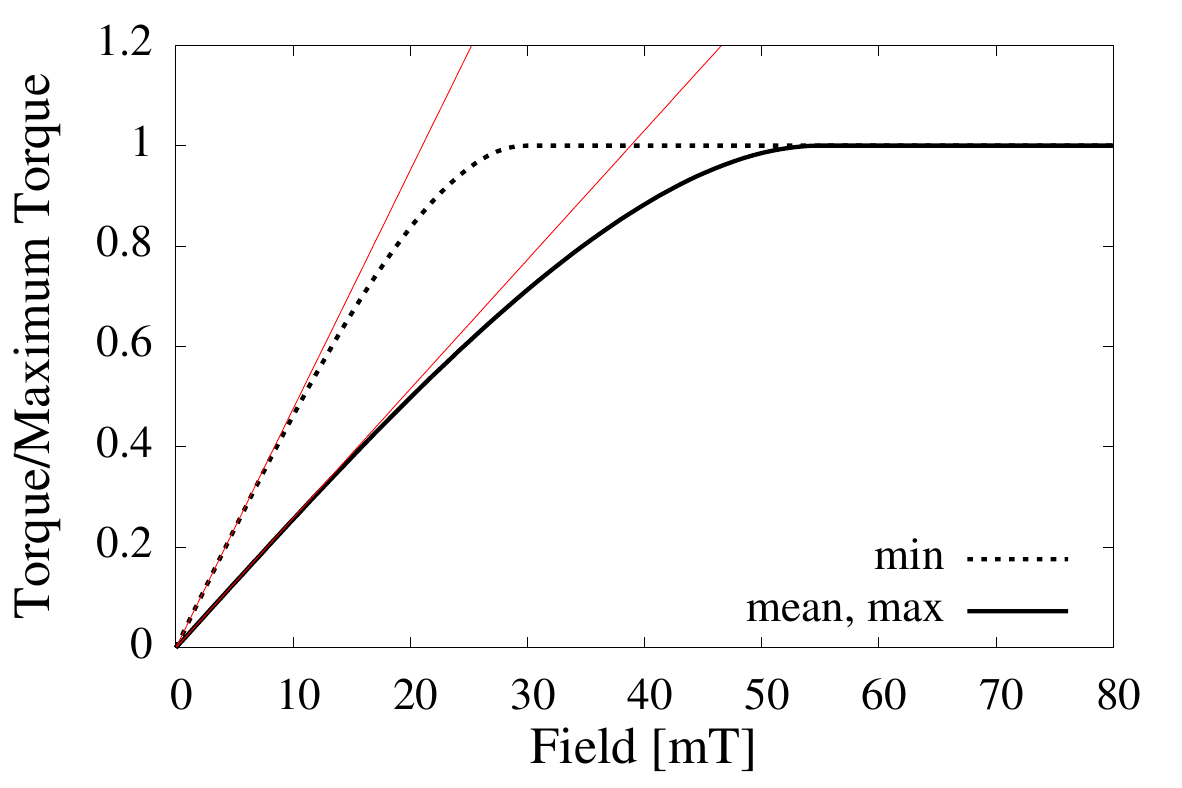}
  \end{centering}
  \caption{Torque on magneto-tactic bacteria, normalized to the
    maximum torque, as a function of applied field for the smallest
    bacterium in table~\ref{tab:MTB_torque} (min) and the largest
    (max). Since $\Delta N$ for an average bacterium is almost equal
    to that of the largest bacterium, its curve overlaps with the
    curve for the largest bacterium. The red solid asymptotes show the
    approximation for $\Gamma_\parallel=\bs{m}\times\bs{B}$.}
\label{fig:MTBTorqueVersusField}
\end{figure}

Based the extreme values of table~\ref{tab:MTB_torque}, we can
estimate the field dependence of the torque of magneto-tactic bacteria
using the theory of section~\ref{sec:theory}. For convenience, the
field dependence of the torque, normalized to the maximum torque
$\Gamma_{\parallel\text{max}}$, is plotted in
figure~\ref{fig:MTBTorqueVersusField}. Also shown is the approximation
for the case when the magnetisation remains aligned with the easy
axes, $\Gamma_\parallel=\bs{m}\times\bs{B}$, which is valid up to
fields of about \SI{10}{mT}.

\section{Conclusion}
\label{sec:conclusion}
We have modelled and measured the magnetic torque and maximum rotation
frequency of microfabricated elements with dimensions in range of 100 to
\SI{500}{\micro m}, as well as magnetotactic bacteria that have a
length of a few \si{\micro m}.

Using torque magnetometry we were able to measure the torque of an
array of 170 microfabricated elements as a function of applied field
angle, but only for fields applied out of the film plane.  The maximum
out-of-plane torque is reached at fields of about \SI{1.2}{T}, and
agrees within measurement errors with our model. For a single
element the out-of-plane torque is in the order of \SI{1}{nNm}. We
calculate the in-plane torque to be in the order of \SI{1}{pNm}, which
is below the sensitivity of our torque magnetometer, even in a large array.

The maximum rotation speed of microfabricated elements and
magnetotactic bacteria under application of a rotating external field
was measured. For microfabricated elements, the maximum rotation
frequency drops with the length of the element. For our \SI{50}{\micro
  m} wide microfabricated elements, we find maximum rotation speeds in
the order \SI{10}{rad/s}, which agrees within error boundaries with
our model based on the competition between magnetic and drag torque
for elements of 250 and \SI{500}{\micro m} length. For elements of
\SI{100}{\micro m} and all \SI{25}{\micro m} wide elements however,
the measured maximum rotation frequency is a factor of three to four
lower than predicted by this model. We suspect that this is caused by
incomplete saturation of the magnetisation in the elements.

By means of finite elements simulations, we show that the maximum
magnetic torque on the magnetosome chain can be accurately
calculated by an analytic \emph{chain of spheres} model. From TEM
images of the magnetosomes, we calculate the average maximum magnetic
torque to be about \SI{1}{aNm}. The values range between 0.02 and
\SI{4.4}{aNm}, due to the large spread in bacteria morphology.

Using average bacteria dimensions measured by SEM, combined with the
maximum torque determined from TEM image, we calculate an average
maximum rotation frequency of \SI{20}{rad/s}, with values ranging from
8 to \SI{66}{rad/s}. The maximum measured angular velocity of
magnetotactic bacteria in a rotating magnetic field is in the order
of \SI{10}{rad/s}, which is in fair agreement with our model.

Summarising, our experiments show that a simple model based on
magnetic torque and viscous drag describes the maximum rotation
velocity for microfabricated elements for two out of six investigated
combinations of outer dimensions and magnetic element sizes within
measurement accuracy, but overestimates this value by at most a factor
of three for the others.  Application of this model to magnetotactic
bacteria leads to predictions of maximum rotation frequencies which
are in agreement with observations. In the first place, this confirms
the assumption that magnetotactic bacteria act passively on the
external magnetic field, like compass needles. Secondly, the theory
developed in this paper will support the analysis of experiments
with magnetic objects in liquid, which is for instance the case in the
field of medical microrobotics.

\begin{acknowledgments}
  The authors wish to acknowledge the contribution of Hans Kolk, who
  did a major part of the measurements during his bachelor study. We
  are grateful for the contribution of Remco Sanders in the
  preparation of magnetotactic bacteria and and follow up
  experiments, Thijs Bolhuis for magnetic characterization and Jasper
  Keuning for setting up and performing initial experiments and
  programming of the image recognition software. This work was
  partially funded by the MIRA Research Institute of the University of
  Twente.
\end{acknowledgments}

%%%%%%%%%%%%%%%%%%%%--------- BIBLIOGRAPHY ---------%%%%%%%%%%%%%%%%%%%%
%\bibliographystyle{apsrev_modified}
%\bibliographystyle{bst/apalikeleon}
\bibliographystyle{bst/apsrev_modified_doi}
\bibliography{paperbase}
%%%%%%%%%%%%%%%%%%%%%%%%%%%%%%%%%%%%%%%%%%%%%%%%%%%%%%%%%%%%%%%%%%%%%%%%%%%%%%
\end{document}

%% file: includes/LeonHeader.tex
\usepackage[pdftex,
        colorlinks=true,
        urlcolor=darkblue,               % \href{...}{...}
        anchorcolor=darkblue,
        filecolor=green,                 % \href*{...}
        linkcolor=darkblue,              % \ref{...} and \pageref{...}
        menucolor=darkblue,
        citecolor=darkblue,
        pagebackref,
        backref={\pdfbackref},
        pdfpagemode={UseOutlines},
        bookmarks=true,
        bookmarksopen=true,
        pdftitle={\pdftitle},
        pdfauthor={\authorname}, 
        pdfsubject={\pdfsubject},
        pdfkeywords={\pdfkeywords},
        ]{hyperref}

\usepackage{graphicx}
\DeclareGraphicsExtensions{.pdf,.png,.jpg}
\usepackage{thumbpdf}
\usepackage{color}
\definecolor{darkgreen}{rgb}{0,0.5,0}
\definecolor{darkblue}{rgb}{0,0,0.5}
\definecolor{brown}{rgb}{0.98,0.92,0.73}
\definecolor{red}{rgb}{1,0,0}
\definecolor{yellow}{rgb}{1,1,0}
\definecolor{blue}{rgb}{0,0,1}
\definecolor{green}{rgb}{0,1,0}
\definecolor{purple}{rgb}{1,0,1}
\definecolor{gray}{rgb}{0.8,0.8,0.8}
\definecolor{black}{rgb}{0,0,0}
\definecolor{white}{rgb}{1,1,1}
\definecolor{gold}{rgb}{1.,0.84,0.}

\graphicspath{{~/Documents/Images//}{Figures//}}

\usepackage{amsmath}
\usepackage{wasysym}
\usepackage{booktabs}

\usepackage{lineno}

\usepackage{pdfsync}

\usepackage{doi}

\usepackage{siunitx}

\usepackage{subfig}

%% file: includes/LeonMacro.tex
\usepackage{amsmath}
\def\bs{\boldsymbol}

\newcommand{\grad}[1]{%
        \bs{\nabla} #1
}
\newcommand{\diver}[1]{%
        \bs{\nabla}\cdot\bs{#1}
}
\newcommand{\rot}[1]{%
        \bs{\nabla}\times\bs{#1}
}